\newcommand{\longversion}{0}
\documentclass[10pt]{article} 
\usepackage[preprint]{tmlr}

\usepackage{amsmath,amsfonts,bm,amsthm}

\def\eqref#1{equation~\ref{#1}}

\def\1{\bm{1}}

\def\eps{{\epsilon}}

\DeclareMathAlphabet{\mathsfit}{\encodingdefault}{\sfdefault}{m}{sl}
\SetMathAlphabet{\mathsfit}{bold}{\encodingdefault}{\sfdefault}{bx}{n}

\RequirePackage[
  datamodel=acmdatamodel,
  style=acmnumeric,
  ]{biblatex}

\usepackage{thm-restate}
\usepackage{multirow}
\usepackage{graphicx}
\usepackage{subfig}
\usepackage{mathtools}

\usepackage{graphicx}
\usepackage{authblk}
\usepackage{xcolor}

\usepackage{algorithmicx,algpseudocode, algorithm}
\usepackage{comment}

\usepackage{cleveref}

\newcommand{\abs}[1]{\lvert{#1}\rvert}

\newcommand{\calN}{\mathcal{N}_{\mathbb{Z}}}

\theoremstyle{plain}
\newtheorem{theorem}{Theorem}
\numberwithin{theorem}{section}
\newtheorem{lemma}[theorem]{Lemma}
\numberwithin{lemma}{section}
\newtheorem{corollary}[lemma]{Corollary}
\numberwithin{corollary}{section}
\newtheorem{definition}{Definition}
\numberwithin{definition}{section}

\numberwithin{assumption}{section}

\numberwithin{example}{section}

\numberwithin{proposition}{section}
\newtheorem{claim}[theorem]{Claim}
\numberwithin{claim}{section}

\makeatletter
\newcommand{\settitle}{\@maketitle}
\makeatother

\begin{document}

\title{Continual Release of Differentially Private Synthetic Data from Longitudinal Data Collections}

\author[1]{Mark Bun}
\author[1]{Marco Gaboardi}
\author[2, 3]{Marcel Neunhoeffer}
\author[4]{Wanrong Zhang}

\affil[1]{Boston University, Boston, USA}
\affil[2]{Institute for Employment Research, Nuremberg, Germany}
\affil[3]{Ludwig-Maximilians-Universit\"at, Munich, Germany}
\affil[4]{Harvard University, Cambridge, USA}

\renewcommand\Affilfont{\itshape\small}





\maketitle

\begin{abstract}
Motivated by privacy concerns in long-term longitudinal studies in medical and social science research, we study the problem of continually releasing differentially private synthetic data from longitudinal data collections. We introduce a model where, in every time step, each individual reports a new data element, and the goal of the synthesizer is to incrementally update a synthetic dataset in a consistent way to capture a rich class of statistical properties. We give continual synthetic data generation algorithms that preserve two basic types of queries: fixed time window queries and cumulative time queries. We show nearly tight upper bounds on the error rates of these algorithms and demonstrate their empirical performance on realistically sized datasets from the U.S. Census Bureau's Survey of Income and Program Participation.
\end{abstract}

\textbf{Please cite as:} Mark Bun, Marco Gaboardi, Marcel Neunhoeffer, and Wanrong Zhang. 2024. Continual Release of Differentially Private Synthetic Data from Longitudinal Data Collections. Proc. ACM Manag. Data 2, 2, Article 94 (May 2024), 26 pages. https://doi.org/10.1145/3651595

  \ifcsname forappendix\endcsname%
    \ifnum\longversion=1
    \renewcommand{\forappendix}[1]{{#1}}
     \else
    \renewcommand{\forappendix}[1]{}
      \fi
  \else%
   \ifnum\longversion=1
    \newcommand{\forappendix}[1]{{#1}}
     \else
    \newcommand{\forappendix}[1]{}
      \fi
  \fi%

      \ifnum\longversion=1
    \newcommand{\forshort}[1]{}
     \else
    \newcommand{\forshort}[1]{{#1}}
      \fi

\section{Introduction}

In a \emph{longitudinal study}, research subjects are repeatedly observed over an extended period of time. Longitudinal studies are essential in medicine and public health, where they are used to establish risk factors for disease (indeed, the term ``risk factor'' itself comes from the influential Framingham Heart Study~\cite{KannelDKRS61}); in developmental psychology to track physical, social, and emotional development through childhood and aging; in business studies to understand business growth and competition and how they affect the credit and labor markets; and in economics to understand employment and income levels and their relation to education, family, and significant life events. Longitudinal designs can be advantageous over single-shot designs in that they provide insight into group and individual-level changes over time. As an example, this aspect of the British Doctors' survey led to the first conclusive evidence of the link between smoking and lung cancer~\cite{DollH56}.

Nevertheless, longitudinal studies come with several important practical challenges, as well as complexity in devising sound statistical methodology. The challenge we address in this work is maintaining the privacy of research subjects while accommodating the workflows and expectations of data analysts. Specifically, we wish to understand when we can release longitudinal studies under the guarantees of \emph{differential privacy}~\cite{DMNS06}. Toward achieving practical desiderata for analyzing longitudinal US Census Bureau data~\cite{VilhuberAR16,BenedettoSA13,BenedettoST18,slbd2023}, we do so by generating differentially private synthetic data from longitudinal studies. We moreover track the structure of the longitudinal Census studies~\cite{sipp2021,lbd2021} by releasing synthetic data updates continuously over time.

Analysts of privacy-protected datasets often want access in the form of synthetic data (a.k.a. synthesized microdata), which are sets of individual records whose statistical properties are similar to those of the original dataset. Synthetic data is a requirement for some releases, e.g., the 2020 Census of Population and Housing tabulations. In general, synthetic data facilitates exploratory data analysis and the use of existing analysis pipelines. These are often the reasons that have motivated their use in longitudinal surveys~\cite{BenedettoST18,slbd2023}. There is a vast literature on differentially private synthetic data, starting with the work of Barak et al.~\cite{BarakCDKMT07} and Blum et al.~\cite{BlumLR08}. Ullman and Vadhan~\cite{Ullman2011} showed that releasing differentially private synthetic is computationally intractable in general, hence, most of the literature has focused on methods that are efficient in practical scenarios~\cite{hardt2010multiplicative,GaboardiAHRW14,10.1145/3134428,VietriTBSW20,McKennaMS21,neunhoeffer2021private,McKennaMSM22,pmlr-v202-liu23ag,wang2023postprocessing}.
These works developed a wide range of techniques for releasing differentially private synthetic data, but none so far have addressed longitudinal studies and releasing them continuously over time. This is what we aim to do here. 

An important feature, specific to longitudinal data collections, is that every individual's data  is collected over multiple reporting periods. Thus, changes to an individual's data over time may be reflected in, and can inform the study of, the population. For example, in longitudinal survey studies such as the US Census Bureau's Survey of Income and Program Participation (SIPP)~\cite{sipp2021}, individuals are surveyed with the same set of questions over a period of time spanning multiple reporting periods. 
Answers to these questions not only summarize how the population changes over time, but also summarize the individual-level trends that appear. For example, tracking responses to the question ``Were you employed this month?'' not only enables estimating population-level unemployment and how that changes over time, but \emph{also} enables monitoring other statistics such as the lengths of unemployment spells.
It is natural to require that differentially private synthetic data from longitudinal data collections preserve this feature. This induces consistency constraints on the way the synthetic data are generated. We will describe these constraints in more detail below, and see the major role they play in our model.

The temporal aspect of longitudinal data collections that we just discussed also makes it natural to conduct studies or release data repeatedly over time. For example, the results of several US Census longitudinal studies are published continuously every year~\cite{sipp2021,lbd2021}.
While most algorithmic research in differential privacy focuses on single-shot data analysis, alternative models are well-equipped to address this aspect of longitudinal data collections.  For instance, in the \emph{continual observation} model of Dwork, Naor, Pitassi, and Rothblum~\cite{DNPR10} and Chan, Shi, and Song~\cite{chan2011private}, individual data is presented to an algorithm in a streaming fashion, and the goal of the algorithm is to produce an estimate of one or more summary statistics of the data at every time step. 
In the continual observation model two notions of privacy have been considered: "event-level" and "user-level". The latter guarantees that an algorithm's entire sequence of outputs is insensitive to changing any individual's contribution, even if that individual's data appears arbitrarily many times in the stream. We will use a similar notion in our model to capture part of the longitudinal requirement. Various problems have been studied in the continual observation model~\cite{MirMNW11, JainKT12, BolotFMNT13, KellarisPXP14, ChenMHM17, SongLMVC18, JainRSS21, CardosoR22, AlabiBC22, EpastoMMMVZ23}, primarily motivated by networking and internet monitoring applications with extremely long time horizons. Here instead we focus on a shorter time horizon with the characteristics we discussed before.

\paragraph{Our model}

We propose the following model to formally capture the concept of \emph{continuously releasing differentially private synthetic data from longitudinal studies}. Let $\mathcal{X}$ be a data universe and let $T$ be a time horizon. In each round $t = 1, \dots, T$, a synthetic data generation algorithm $\mathcal{A}$ is given a vector of updates $D^t = (x_1^t, \dots, x_n^t) \in \mathcal{X}^n$, consisting of one update from each of $n$ individuals, and is required to produce a synthetic data vector $\hat{D}^t = (\hat{x}_1^t, \dots, \hat{x}_m^t) \in \mathcal{X}^m$, consisting of one update for each of $m$ \emph{synthetic individuals}. We consider $m$ and $n$ fixed over the whole time horizon. The size $m$ of the synthetic output may differ from the size $n$ of the input data, as often happens in traditional differentially private synthetic data generation. 

The sequence of synthetic data releases is generated with respect to a pre-specified class of queries
$\mathcal{Q} = \cup_{t = 1}^T \mathcal{Q}_t = \cup_{t = 1}^T \{q: (\mathcal{X}^t)^n \to \mathbb{R}\}$ that is segmented according to time. Abusing notation slightly, we'll assume that the query segments are nested $\mathcal{Q}_t \subseteq \mathcal{Q}_{t+1}$ in that for every $q \in \mathcal{Q}_t$, the query $q'(D^1, \dots, D^{t+1}) := q(D^1, \dots, D^t)$ is in $\mathcal{Q}_{t+1}$. Notice  that results of queries may depend on the whole individuals' history up to time $t$.

The algorithm $\mathcal{A}$ should satisfy the following two requirements:

\paragraph{Differential privacy.} $\mathcal{A}$ should be differentially private with respect to changing any individual's entire sequence of updates $(x_i^1, \dots, x_i^T)$. This requirement is analogous to user-level privacy in the continual observation model. Our results are stated for a static (non-adaptive) adversary that selects neighboring datasets at the beginning of the experiment but which are only revealed to the DP algorithm incrementally. This is the original privacy model considered in the early continual release papers \cite{DNPR10, chan2011private}.

\paragraph{Accuracy with respect to $\mathcal{Q}$.} For every $t = 1, \dots, T$ and every query $q \in \mathcal{Q}_t$, it should hold that
$q(\hat{D}^1, \dots, \hat{D}^t) \approx  q(D^1, \dots, D^t).$
That is, the synthetic dataset produced up to time $t$ should be accurate with respect to the set of queries that are well-defined up to time $t$. \\

We remark once again that the accuracy requirement, combined with the fact that queries can explore the whole individual's history up to time $t$, imposes  \emph{consistency} constraints on the data of individuals across the different synthetic data releases. This is necessary to capture the individual-level trends which are often essential to longitudinal studies, as discussed above.

To see what can go wrong without explicitly incorporating consistency constraints, consider the following first attempt at continually generating private synthetic data from longitudinal data collections: Simply recompute a new synthetic dataset from scratch in every round. That is, in each time step $t$, one could apply a single-shot synthetic data generator to the portion of the dataset observed up to time $t$ that yields accurate answers with respect to $\mathcal{Q}_t$ (and, by our assumption that queries are nested, to $\mathcal{Q}_{t'}$ for every $t' < t$ as well). Composition theorems for differential privacy show that the worst-case error incurred overall is roughly $\sqrt{T}$ times larger than the single-shot generator's. But in addition to this automatic hit to accuracy, the synthetic dataset produced at time $t+1$ may consist of entirely different records than that at time $t$. 
This can create inconsistency between analyses conducted on the synthetic data after each of these time steps, especially when studying individual-level trends. For instance, without respecting consistency, it may be possible for the number of synthetic individuals who have ever experienced a 6-month unemployment spell to \emph{decrease} from time step $t$ to $t+1$.

In contrast, our problem specification requires synthetic individuals to persist over time and for the longitudinal synthetic data generator to update their records incrementally. This is challenging, as even at time step $t=1$, the synthetic data generator must prepare synthetic records $(\hat{x}_1^t, \dots, \hat{x}_m^t)$ in anticipation of all future queries and arbitrary future data records, as it cannot go back and change these values once they are released. 
Our consistency requirement is not only natural from the standpoint of using synthetic data from longitudinal data collections but as we shall see below, this way of thinking about the problem helps us design algorithms whose accuracy does not need to incur a $\operatorname{poly}(T)$ overhead over the single-shot case.

\paragraph{Our results.} We study continual
differentially private synthetic data generation from longitudinal studies for two fundamental classes of queries. We give theoretical error guarantees essentially matched by lower bounds. Moreover, we apply our algorithms on several Census data examples to illustrate how these theoretical guarantees translate to reasonable accuracy levels for social science datasets. A motivating case study for the problems in this paper is the US Census Bureau's Survey of Income and Program Participation (SIPP)~\cite{sipp2021}. The SIPP conducts monthly interviews of individuals to track income, employment, family composition, and participation in food and unemployment assistance programs. The problem of generating synthetic data for SIPP was previously considered in~\cite{BenedettoSA13, BenedettoST18}, but without the formal protections of differential privacy.

The two classes of queries we consider are part of the bigger class of \emph{counting queries}, which are determined by a predicate $q : \mathcal{X}^t \to \{0, 1\}$ and ask what fraction of the dataset satisfy $q$. That is, such a predicate extends to a query defined over entire datasets $D = (D^1, \dots, D^t)$ where $D^i \in \mathcal{X}^n$ by taking $q(D^1, \dots, D^t) = \frac{1}{n} \sum_{i = 1}^n q(x_i^1, \dots, x_i^t)$. For simplicity and clarity, we take $\mathcal{X} = \{0, 1\}$ for the classes of queries we consider. That is, each individual contributes a new bit of data in each reporting period. The solutions we develop for fixed time window queries (described below) naturally extend to handle categorical data with more than $2$ categories. 

\paragraph{Fixed time window queries.} Suppose we wish to explore the incidence of various short-term trends in the data. For instance, for some parameter $k$, we may be interested in understanding what fraction of the population was below the poverty line in each consecutive window of $k$ months and tracking how this incidence changes in response to historical events (such as broader economic downturns). We study the class of queries for which accuracy captures the following condition: When restricted to every time window of width $k$ (for every $t\geq k$), the histogram of the synthetic data approximates the corresponding histogram on the original dataset. That is, for every pattern $s \in \{0, 1\}^k$ and every time step $t$, the fraction of synthetic individuals for whom $(\hat{x}_i^{t-k+1}, \dots, \hat{x}_i^t) = s$ approximates the fraction of original dataset members for whom $(x_i^{t-k+1}, \dots, x_i^{t}) = s$.

We give an algorithm for this problem that guarantees concentrated differential privacy (CDP)~\cite{dwork2016concentrated, bun2016concentrated}, such that the worst-case error for any histogram bin is $\widetilde{O}(2^k\sqrt{T} / n)$, which improves to $\widetilde{O}(\sqrt{kT} / n)$ for ``average'' bins containing at most an $O(1/2^k)$ fraction of the dataset. See Corollary~\ref{cor:rel-error} for the exact error guarantee.
Furthermore, we give a simple method by which an analyst can postprocess the generated synthetic data to obtain estimates with significantly less bias and error $\widetilde{O}(\sqrt{kT} / n)$ for all bins. For constant $k$, these results essentially match a lower bound of $\Omega(\sqrt{T}/n)$ that holds even for the easier single-shot problem of releasing accurate answers with respect to $\mathcal{Q}_T$ when the entire dataset is revealed~\cite{bun2016concentrated}. Moreover, accuracy for histogram bins immediately implies accuracy for arbitrary queries defined over contiguous time windows of length $k$. If a query $q$ can be written as a linear combination $q(x^{t-k+1}, \dots, x^{t}) = \sum_{s \in \{0, 1\}^k} w_s \mathbb{I}((x^{t-k+1}, \dots, x^t) = s)$, then the synthetic data has error $\tilde{O}(2^k\|w\|_2\sqrt{T}/n)$ with respect to $q$.

\paragraph{Cumulative time queries.}

In addition to understanding phenomena over fixed time windows, we may also be interested in tracking longer-term trends over individual's bitstrings. When $\mathcal{X} = \{0, 1\}$, a natural statistic to keep track of is the Hamming weight (cumulative number of 1's) of individuals' histories. This, for instance, enables tracking, for every time step $t$ and every value $b$, the fraction of individuals who have been below the poverty line for at least $b$ out of the first $t$ months. We give a CDP algorithm for this problem with worst-case error $\tilde{O}(\sqrt{T}/n)$, which nearly matches a lower bound of $\Omega(\sqrt{T}/n)$ that again holds even for the final single-shot problem.\\

Our algorithms for both problems follow a natural two-stage approach. For each time step $t$, in the first stage, we privately compute noisy estimates for the new queries in $\mathcal{Q}_t$. In the second stage, we identify updates to the synthetic data that make it approximately consistent with these noisy statistics. For both problems we consider, the first stage is straightforward from standard techniques from the differential privacy literature: Independent noise addition to producing noisy histograms in the case of fixed time window queries and an adaptation of the classic \emph{tree-based aggregation mechanism}~\cite{DNPR10, chan2011private} for cumulative time queries.

The second stage, in each case, requires additional ideas. For the fixed time window problem, our ability to generate synthetic bits that will fit the new noisy statistics requires us to post-process these statistics so that they meet the following consistency constraints: For every $z \in \{0, 1\}^{k-1}$, we have $p_{z0}^t + p_{z1}^t = p_{0z}^{t-1} + p_{1z}^{t-1}$, where $p_s^t$ denotes the number of synthetic records that end in the substring $s$ at time $t$. That is, the number of synthetic records ending with the suffix $z$ at time $t-1$ must be identical to the number of synthetic records ending with either $z0$ or $z1$ at time $t$. 

We solve this set of consistency constraints such that the error on each histogram bin can be written as a certain linear combination of errors of the noisy statistics generated in stage 1. We can then leverage the independence of these errors to give a sharp bound on the post-processed errors. 

Another issue in synthetic data generation is the possibility that noisy histogram bin counts may go negative.
Na\"ively clamping counts to be nonnegative does not work, as we cannot resurrect nonzero counts from zero counts at subsequent time steps. We overcome this issue by padding the initial synthetic dataset so that every histogram bin is well-represented. With high probability over the entire run of the algorithm, no noisy counts will ever go so negative to cancel out the padding completely. Note that while the padding introduces bias into the estimates of each histogram bin, the amount of this bias is public information and can be corrected in additional post-processing if the analyst chooses to do so.

Generating the next round of synthetic bits is somewhat simpler for cumulative time queries. But its feasibility relies on the statistics computed in stage 1 to satisfy the following two monotonicity constraints:  For every value of $b$ and $t$, the estimated count of individuals whose data has Hamming weight at least $b$ by time $t$ must be (1) at least the estimate produced for weight $\ge b$ at time step $t-1$ and (2) at most the estimate produced for weight $\ge b-1$ at time step $t-1$. The first condition corresponds to the fact that (synthetic) individuals' Hamming weights can only increase over time. The second corresponds to the fact that Hamming weights can only increase by at most $1$ each time step. Neither condition is automatically guaranteed by the tree-based aggregation mechanism but can be ensured by an additional monotonization procedure that adapts a technique from~\cite{chan2011private} for ensuring the monotonicity condition (1).

\subsection{Related work} \label{sec:related}

While, to our knowledge, the problem of continuously
releasing differentially private synthetic data from longitudinal studies has yet to be studied, solutions to related questions have previously appeared in the differential privacy literature. Motivated by the problem of counting distinct elements in data streams, Ghazi, Kumar, Manurangsi, and Nelson~\cite{ghazi2022private} studied variants of our cumulative time queries problem. In their terminology, our cumulative time queries problem corresponds to answering $\mathrm{CountOcc}^{\ge b}$ queries, for all $b$ simultaneously, in their cumulative, item-level DP, bundle setting. Our solution also immediately implies an $\tilde{O}(\sqrt{T}/n)$-error algorithm for answering $\mathrm{CountOcc}^{=b}$ queries for all $b$ simultaneously in their time window, item-level DP, bundle setting that follows from the fact that the number of individuals producing bit $1$ exactly $b$ times between time steps $t_1 < t_2$ can be expressed as the difference between the number of individuals with Hamming weight at least $b$ up to time $t_2$ and the number of individuals with Hamming weight at least $b-1$ up to time $t_1$. Using similar constructions to ours (and also building on a prior reduction of Bolot, Fawaz, Muthukrishnan, Nikolov, and Taft~\cite{BolotFMNT13}), Ghazi, Kumar, Manurangsi, and Nelson~\cite{ghazi2022private} studied these and related problems when the goal is to answer queries for any fixed single value of $b$. Our algorithm simultaneously permits accurate estimation of these queries for all $b$ while also incrementally generating synthetic data matching these query answers. 

Several works have studied how to privately release statistics of longitudinal data collections under the stronger constraint of \emph{local differential privacy}. Google's RAPPOR system~\cite{ErlingssonPK14} enables tracking the population-level average of a single Boolean statistic (corresponding to our fixed time windows problem with $k = 1$) under a heuristic assumption that each individual's bit flips its value only a small number of times.
These ideas were developed further by Erlingsson, Feldman, Mironov, Raghunathan, Talwar, and Thakurta~\cite{erlingsson2019amplification}, who gave an algorithm with error that scales with an upper bound on the number of times any individual's bit can flip its value rather than the overall time horizon. And while it is not posed as a synthetic data generation problem, the task of incrementally generating synthetic data from their noisy estimates is immediate when $k = 1$. Joseph, Roth, Ullman, and Waggoner~\cite{joseph2018local} studied a similar problem, but in a model where it is further assumed that individuals come from distinct subpopulations, and in each time step, each individual independently draws a fresh bit from their subpopulation-specific distribution. Additional work on continual frequency estimation in the local model includes~\cite{DingKY17, OhrimenkoWW22, XueYHZW22, arcolezi2022frequency}.

For simplicity, we implement our algorithm for synthetic data release for cumulative time queries using the classic binary tree counter in Section \ref{sec.exp}. However it could be implemented using an arbitrary differentially private algorithm for tracking the sum of a stream of bits. Our formal description of Algorithm~\ref{algo:cumulative} is indeed stated in terms of a generic stream counter algorithm. Stream counters enjoying improved concrete accuracy guarantees have been the focus of recent attention~\cite{Honaker15, DenisovMRST22, FichtenbergerHU22, HenzingerUU23}, and using them in place of the tree counter in our work may yield improved practical results.

\section{Preliminaries}

\subsection{Continual synthetic data}\label{sec:model}

Let $\mathcal{X}$ be a data universe and $T$ be a known time horizon. The data of an individual $x_i = (x_i^1, \dots, x_i^T)$ is a sequence of universe items where each $x_i^t \in \mathcal{X}$ is the report arriving at time $t$. For a dataset of $n$ individuals, let $D^t = (x_1^t, \dots, x_n^t)$ be the vector of data reported at time $t$. The entire dataset may thus be written as $D = (D^1, \dots, D^T)$. For simplicity throughout the rest of this paper, we will take $\mathcal{X} = \{0, 1\}$ unless otherwise stated. We are interested in releasing synthetic data that is accurate with respect to a class of queries $\mathcal{Q} = \cup_{t = 1}^T \mathcal{Q}_t$, where each $\mathcal{Q}_t$ is a set of queries of the form $q : (\mathcal{X}^t)^* \to \mathbb{R}$ that are defined on prefixes of length $t$.

\paragraph{Continual release of synthetic data.} The goal of a synthetic data generation algorithm is as follows. In each round $t$, it receives as input a vector of reports $D^t = (x_1^t, \dots, x_n^t)$. It then produces a synthetic vector of reports $\hat{D}^t = (\hat{x}_1^t, \dots, \hat{x}_m^t)$. For parameters $\alpha, \beta \in (0, 1)$, we say that the synthetic data generation algorithm is $(\alpha, \beta)$-accurate with respect to $\mathcal{Q} = \cup_{t = 1}^T \mathcal{Q}_t$ if, for every input dataset $D$, with probability at least $1-\beta$ over the coin tosses of the algorithm, it holds that for every $t = 1, \dots, T$ and $q \in \mathcal{Q}_t$, $|q(D^1, \dots, D^t) - q(\hat{D}^1, \dots, \hat{D}^t)| \le \alpha.$

\medskip

In this work, we focus on releasing synthetic data for two basic classes of counting queries. A counting query is specified by a predicate $q : \mathcal{X}^t \to \{0, 1\}$, for some $1 \le t \le T$, and extends to an entire dataset by averaging: $q(D^1, \dots, D^t) = \frac{1}{n} \sum_{i = 1}^n q(x_i^1, \dots, x_i^t)$. The classes of queries we study are:

\paragraph{Fixed time window queries.} Fix a parameter $k \in \{1, \dots, T\}$. For each string $s \in \{0, 1\}^k$, and each $t = k, k+1, \dots, T$, define the query $q_s^t(x^1, \dots, x^t) = \mathbb{I}((x^{t-k+1}, \dots, x^t) = s)$. The query $q_s^t$ indicates whether the length-$k$ suffix of data item $x$ that has arrived by time $t$ equals $s$. For each $t$, the set of queries $\{q_s^t\}_{s \in \{0, 1\}^k}$ jointly capture this histogram of $k$-bit suffixes of the portion of the data that has arrived by time $t$.

\paragraph{Cumulative time queries.}
For each $b = 0, 1, \dots, T$ and each $t = 1, \dots, T$, define the query $c_b^t(x^1, \dots, x^t) = \mathbb{I}(x^1 + \dots + x^t \ge b)$. That is, $c_b^t$ indicates whether the portion of data item $x$ that has arrived by time $t$ has Hamming weight at least $b$.

\paragraph{Reducing cumulative time queries to fixed time window queries.} If we set $k = T$, observe that each query $c_b^t$ may be written as $c_b^t(x) = \sum_{s \in \{0, 1\}^k : |s| \geq b} q_s^t(x)$. (Here, we are adopting the convention that each $x_i^t = 0$ for $t \le 0$.) This implies that an accurate synthetic data generator for fixed time window queries with $k = T$ implies one for the cumulative time queries problem with a roughly $2^k$-factor blowup in error. While this shows that the problems are related, our algorithm which is tailored to cumulative time queries achieves significantly better accuracy.

\subsection{Differential privacy background}

Differential privacy is a mathematical notion of privacy for statistical data analysis which bounds the amount of information an adversary can learn about any individual. Given a space of datasets $\mathcal{X}^T$, we say that two datasets $D, D'$ are {\em neighboring} if they differ in one individual's information. 
\forappendix{
\begin{definition}[Differential Privacy \cite{DMNS06, dwork2006our}]\label{def.dp}
    A randomized algorithm $\mathcal{M}: \mathcal{X}^T \rightarrow \mathcal{R}$ is \emph{$(\varepsilon,\delta)$-differentially private} if for every pair of neighboring datasets $D, D'\in \mathcal{X}^T$, and for every subset of possible outputs $\mathcal{S} \subseteq \mathcal{R}$,
    \begin{equation*}
        \Pr[\mathcal{M}(D) \in \mathcal{S}] \leq e^\varepsilon\cdot \Pr[\mathcal{M}(D') \in \mathcal{S}] + \delta.
    \end{equation*}
\end{definition}
}
{\em Zero-concentrated differential privacy (zCDP)}~\cite{dwork2016concentrated,bun2016concentrated} is a variant of differential privacy that quantifies the closeness of distributions differently, and is especially amenable to analyzing Gaussian noise addition.

\begin{definition}[Zero-Concentrated Differential Privacy (zCDP) \cite{bun2016concentrated}]
    A randomized algorithm $\mathcal{M}: \mathcal{X}^T\rightarrow \mathcal{R}$ is \emph{$\rho$-zCDP} if for every pair of neighboring datasets $D, D'\in \mathcal{X}^T$, and for all $\alpha\in(1,\infty)$, $\mathrm{D}_\alpha(\mathcal{M}(D) || \mathcal{M}(D'))\le \rho \alpha,$
 where $\mathrm{D}_\alpha$ denotes the R\'enyi divergence of order $\alpha$.
\end{definition}

\forappendix{
A $\rho$-zCDP guarantee can be translated to a $(\eps, \delta)$-DP guarantee using the following result from \cite{bun2016concentrated}: If $\mathcal{M}$ provides $\rho$-zCDP, then $\mathcal{M}$ is $(\rho + 2\sqrt{\rho \log(\frac{1}{\delta})}, \delta)$-differentially private for every $\delta > 0$.
}
Zero-concentrated differential privacy is also desirable for its straightforward and tight {\em composition}, which characterizes how privacy degrades as performing more computations on the data. 

\begin{theorem}[zCDP Composition]
Let $\mathcal{M}_1: \mathcal{X}^T \rightarrow \mathcal{R}$ is $\rho$-zCDP and $\mathcal{M}_2: \mathcal{X}^T \rightarrow \mathcal{R}$ is $\rho'$-zCDP, then the mechanism defined as $(\mathcal{M}_1, \mathcal{M}_2)$ satisfies $(\rho+\rho')$-zCDP.
\end{theorem}

The {\em discrete Gaussian mechanism}~\cite{canonne2020discrete}  with parameter $\sigma^2$ takes in a function $q$, dataset $D$, and outputs $q(D)+\calN(0,\sigma^2)$. The noise scale is fully specified as $\sigma^2=\frac{\Delta q}{2\rho}$, given the privacy parameter $\rho$ and the query sensitivity $\Delta q$. The variance of $\calN(0, \sigma^2)$ is at most $\sigma^2$.

\begin{definition}[Discrete Gaussian \cite{canonne2020discrete}]
The discrete Gaussian distribution with mean $0$ and scale $\sigma$ is denoted $\calN(0,\sigma^2)$. It is a probability distribution supported on the integers and defined by 
$
    \Pr[X=x]=\frac{\exp(-\frac{x^2}{2\sigma^2})}{\sum_{y\in \mathbb{Z}}\exp(-\frac{y^2}{2\sigma^2}) }. 
$
\end{definition}

\section{Generating Synthetic Data for Fixed Time Window Queries}\label{sec:fixtime}

In this section, we consider fixed time window queries. The goal is to produce differentially private synthetic data such that the histogram of the synthetic data approximates the corresponding histogram on the original dataset for every time window of width $k$. 

\subsection{Algorithm}\label{sec:algo1}

\begin{algorithm}[ht]
\caption{Private synthetic data generation preserving fixed time window queries}\label{algo:fixedwindow}
\begin{algorithmic}

  \State \textbf{Input:} Dataset consisting of $n$ users/streams $(D^1, D^2, \ldots, D^T)$ where $D^t=(x_1^t, \ldots, x_n^t)$ is the vector related to the data produced at time $t$,
  a known time horizon $T$, the time window length $k$, privacy parameter $\rho$, the number of padding records per bin $n_{\text{pad}}$.
  
 \State \textbf{Output:} A starting dataset $\hat{D}^k$ of $n^*$ people with $k$ columns. Update vectors $\hat{D}^{t}=(\hat{x}_1^{t}, \ldots, \hat{x}_{n^*}^{t})$ for all steps $t>k$.

\For{ $t \gets k$ to $T$}
\State Calculate the count of individuals that have suffix $(x_i^{t-k+1}, \dots, x_i^{t})$ equal to $s$ as $C_s^t$. 

\State Construct a DP histogram consisting of $2^k$ bins, where the value of each bin corresponding to \State \hspace{2mm} $s\in \{0,1\}^k$ is the noisy count of individuals that have suffix $(x_i^{t-k+1}, \dots, x_i^{t})$ equal to $s$ as \State \hspace{2mm}  $\hat{C}^t_s=C^t_s+n_{\text{pad}}+\calN(0, \frac{T-k+1}{2\rho }).$

\If{$t=k$}

\State Set $p_s^t=\hat{C}^t_s$ for every $s\in \{0,1\}^k$.

\State Output any dataset $D^k$
such that the number of people 
\State \hspace{2mm} that have string $(x_i^{1}, \dots, x_i^{k})$ equal  to $s$ is equal to $\hat{C}^t_s$.

\Else


\For{ every $z \in \{0,1\}^{k-1}.$}

\State Set $\Delta_{z}=\frac{1}{2}( p^{t-1}_{0z}+p^{t-1}_{1z}-(\hat{C}_{z0}^t+\hat{C}_{z1}^t))$

\If{$\Delta_{z}$ is not an integer}

\State Sample $b_z \in \{-1/2, 1/2\}$ uniformly at random.

\State Set $p^{t}_{z0}=\hat{C}_{z0}^t+\Delta_{z}+b_z$ and $p^{t}_{z1}=\hat{C}_{z1}^t+\Delta_{z}-b_z$.

\Else

\State Set $p^{t}_{z0}=\hat{C}_{z0}^t+\Delta_{z}$ and $p^{t}_{z1}=\hat{C}_{z1}^t+\Delta_{z}$.

\EndIf


\State Select $p^{t}_{z1}$ indices, denoted by $I_{z1}$ from the index set \State \hspace{2mm} $I_z = \{i \mid (x_i^{t-k+1}, \ldots, x_i^{t-1}) = z \}$. 


\State Update $\hat{x}_i^{t}\gets 1$ for $i \in I_z$ and $\hat{x}_i^{t} \gets 0$ if $i \in I_z \setminus I_{z1}$.   
\EndFor

\State Output $\hat{D}^{t}$.

\EndIf

\EndFor
  
\end{algorithmic}
\end{algorithm}

This section presents our algorithm (Algorithm~\ref{algo:fixedwindow}) for generating differentially private synthetic data that preserves the histogram over every fixed window of length $k$. Our algorithm consists of two phases at each step. The first phase is to compute the noisy private statistics about the data in the most recent length-$k$ window. The second phase is to construct synthetic data according to the noisy statistics. We elaborate on the details of each phase below.

\paragraph{Producing noisy statistics.}

At each update step of the algorithm, i.e., time steps $t=k, k+1, \ldots, T$, we represent $(D^{t-k+1}, \dots, D^{t})$
as a histogram
with a bin for each string of length $k$, and the value of each bin is the number of appearances of that string in this segment of the dataset. Our algorithm eventually runs for a total of $T-k+1$ update steps. So to privatize the statistics, we construct a DP histogram by independently adding discrete Gaussian noise to each bin: $\hat{C}^t_s=C^t_s+n_{\text{pad}}+\calN(0, \frac{T-k+1}{2\rho })$, where $n_{\text{pad}}$ is a padding parameter we will discuss later.

\paragraph{Generating synthetic data.}
At time $t=k$, we initialize the synthetic dataset to be any dataset such that its histogram representation is consistent with the DP histogram at time $t=k$. 
For $t>k$, when we move a sliding window from time $t$ to time $t+1$, the two windows overlap on a segment of length $k-1$. Given any overlapping segment $z\in \{0,1\}^{k-1}$, synthetic data ending in suffix $z0$ or $z1$ at time $t+1$ must arise by extending synthetic data ending in suffix $0z$ or $1z$ at the previous time $t$.
That is, the synthetic data we generate to approximately fit the new noisy counts $\hat{C}_s$ must be consistent with this overlapping region. More precisely, let $p^t_{s}$ denote the number of appearances of suffix $s$ in the synthetic data in the window of time $t$. Then the synthetic data histogram at $t+1$ must satisfy the consistency constraint that $p^t_{0z}+p^t_{1z} = p^{t+1}_{z0} + p^{t+1}_{z1}$.
To enforce this constraint,
we introduce a correction term $\Delta_{z}=\frac{1}{2}( p^t_{0z}+p^t_{1z}-(\hat{C}_{z0}^{t+1}+\hat{C}_{z1}^{t+1}))$, which depends on the synthetic data histogram at time $t$, as well as the noisy DP histogram at time $t+1$. We set the target synthetic data counts to 
\begin{eqnarray}
p^{t+1}_{z0} & =\hat{C}_{z0}^{t+1}+\Delta_{z} \label{eq.z0}\\
p^{t+1}_{z1} & =\hat{C}_{z1}^{t+1}+\Delta_{z}. \label{eq.z1}
\end{eqnarray}
To generate synthetic data, the algorithm then extends $p^{t+1}_{z0}$ records that ended in $z$ at time $t$ by the bit $0$, and similarly, extends $p^{t+1}_{z1}$ records ending in $z$ by $1$.
Note that the correction term $\Delta_{z}$ may be a half-integer. If this is the case, we set 
\begin{eqnarray}
p^{t+1}_{z0} & =\hat{C}_{z0}^{t+1}+\Delta_{z}+b_z \label{eq.z0}\\
p^{t+1}_{z1} & =\hat{C}_{z1}^{t+1}+\Delta_{z}-b_z, \label{eq.z1}
\end{eqnarray}
where $b_z$ is a rounding term which takes value $-\frac{1}{2}$ with probability $\frac{1}{2}$, and $\frac{1}{2}$ with probability $\frac{1}{2}$. 

We remark that it is possible to view one step of synthetic data generation in each of our algorithms as projecting onto the space of valid synthetic datasets, which can be formulated as an optimization problem: Minimize the distance between the privatized query answers and the consistent query answers while respecting all the consistency constraints. This bears conceptual similarity to classic projection techniques in differential privacy~\cite{NikolovTZ13, HayRMS10}, with the key difference being that those algorithms can exploit the fact that all of the data appears in a single batch, while our algorithm needs to perform projections incrementally. 

Note also that adding noise could result in negative numbers in the DP histograms, which is invalid for generating synthetic data. One possible way to address it is clamping the noisy counts to be non-negative, but this will break the consistency guarantee when continually releasing the synthetic data. 

Instead, we introduce a padding technique to avoid the possibility of negative noisy counts by adding $n_{\text{pad}}$ ``fake'' people to each histogram bin at the very beginning. 
The appropriate choice of $n_{\text{pad}}$ depends on the privacy parameter, time horizon, and the target error probability $\beta_{\text{target}}$. 
Intuitively, when adding $n_{\mathrm{pad}}$ ``fake'' people in each bin, a negative count still occurs if the noise random variable $Y_i$ goes below $-n_{\mathrm{pad}}$. This small error probability can be controlled by estimating the tail of the noise distribution.
Since we are roughly adding discrete Gaussian noise with variance $\sigma^2=\frac{T-k+1}{2\rho}$, we consider the maximal deviation of the independent noise draws over the $2^k(T-k+1)$ bins. 
We have $ \Pr(\min_i Y_i \le -n_{\text{pad}})\le 2^k(T-k+1)\exp\left(-\frac{n_{\text{pad}}\rho}{T-k+1}\right).$
Taking $n_{\text{pad}}=\sqrt{\frac{T-k+1}{\rho}\log \frac{2^k(T-k+1)}{\beta_{\text{target}}}}$ ensures that with probability at least $1-\beta_{\text{target}}$, the DP histograms for all $T-k+1$ update steps are all non-negative. Note that the more careful analysis of our algorithm also needs to account for the correction terms $\Delta$ and rounding terms $b$.

\subsection{Theoretical guarantees}\label{sec:gurantee1}
We now provide our theoretical analysis for Algorithm~\ref{algo:fixedwindow}. We first argue its privacy and then describe the accuracy guarantee which measures the additive error of the fraction of every substring within a fixed window in the synthetic data.

\paragraph{Privacy.} 
Privacy of Algorithm \ref{algo:fixedwindow} follows from composition of the noise-adding mechanism. 
The sensitivity of the count $C_s^t$ is $1$. 
Using the discrete Gaussian mechanism guarantees $\left(\frac{\rho}{T-k+1}\right)$-zCDP per update step. By composition for zCDP, we have $\rho$-zCDP for the entire algorithm.

\begin{theorem}
\label{thm.priv2}
Algorithm \ref{algo:fixedwindow} satisfies $\rho$-zCDP.
\end{theorem}

\paragraph{Accuracy.} 
For each $z \in \{0, 1\}^{k-1}$ and $c \in \{0, 1\}$, denote the true count ending in $zc$ in the original data at time $t$ be $C^{t}_{zc}$.
We are interested in controlling the additive deviation $\abs{\hat{p}^t_{zc}-(C^{t}_{zc}+n_{\text{pad}})}$ between the count of any given suffix in the synthetic data and the padded count from the original data. Perhaps surprisingly, we can maintain a time-uniform bound on the error with high probability. At a high level, we carefully characterize the exact noise propagation and show that the errors are mean $0$ with the same variance over time. Then we can apply a Gaussian tail bound to analyze the error per histogram bin. An additional $\log(2^k(T-k+1))$ for the worst-case error comes from a union bound over at most $2^k(T-k+1)$ bins.

\begin{theorem}\label{thm.acc2}
Let $C_s^t$ be the true count of data records ending in $s$ at time step $t$, and $p_s^t$ be the count in the synthetic data obtained by Algorithm \ref{algo:fixedwindow}.
For every $\beta \in (0, 1)$, with probability at least $1-\beta$, the maximum additive error is bounded by 
\begin{equation}\label{eq.errorfixed}
\max_{s, t} \abs{p^t_{s}-(C^{t}_{s}+n_{\mathrm{pad}})}\le \left(\sqrt{\frac{T-k+1}{\rho}}+\frac{1}{\sqrt{2}}\right)\sqrt{\log\left(\frac{2^k(T-k+1)}{\beta}\right)}.
\end{equation}
\end{theorem}

In particular, as long as $n_{\textrm{pad}}$ is at least the expression on the right-hand-side of~\eqref{eq.errorfixed}, the noisy counts $p_s^t$ will all be non-negative and  the algorithm will succeed at producing synthetic data.

\begin{proof}[Proof of Theorem~\ref{thm.acc2}]

Let us briefly recall our quantities of interest and various intermediate quantities we generate along the way:
\begin{enumerate}
    \item $C_s^t$ is the true count of data records ending in $s$ at time step $t$.
    \item $\overline{C}_s^t = C_s^t + n_{\mathrm{pad}}$ is the true count on the padded dataset.
    \item $\hat{C}_s^t$ is the noisy count arising from the DP noisy histogram algorithm applied to the padded dataset.
    \item $p_s^t$ is the corrected estimated count obtained using Equations~\ref{eq.z0} and~\ref{eq.z1}.
\end{enumerate}

To make the accuracy analysis easier to follow, we denote the total number of update steps of Algorithm \ref{algo:fixedwindow} by $R = T-k+1$. Let us also index update steps by $r = 0, 1, \dots, R-1$, where update step $r$ occurs at time step $t = r + k$.

Our algorithm uses the update rule stated in Equations~\ref{eq.z0} and~\ref{eq.z1}, where $\Delta_{z}=\frac{1}{2}( p^r_{0z}+p^r_{1z}-(\hat{C}_{z0}^{r+1}+\hat{C}_{z1}^{r+1}))$, for any $r \geq 1$. This is equivalent to

\begin{eqnarray}\label{eq.prop}
p^{r+1}_{z0}&=\frac{1}{2}\hat{C}_{z0}^{r+1}+\frac{1}{2}(p^r_{0z}+p^r_{1z}-\hat{C}_{z1}^{r+1}) + b_z \cdot e_z \\
p^{r+1}_{z1}&=\frac{1}{2}\hat{C}_{z1}^{r+1}+\frac{1}{2}(p^r_{0z}+p^r_{1z}-\hat{C}_{z0}^{r+1}) - b_z \cdot e_z,
\end{eqnarray}

where $b_z\in\{-\frac{1}{2}, \frac{1}{2}\}$ is a random rounding term, and $e_z=\mathbf{1}[\Delta_z \text{ is not an integer}]$. Recalling that $\bar{C}_s^r=C_s^r+n_{\text{pad}}$ and using the identity $\overline{C}_{0z}^r + \overline{C}_{1z}^r = \overline{C}_{z0}^{r+1} + \overline{C}_{z1}^{r+1}$
we can decompose the error as follows.
\begin{equation} \label{eqn:error-decomp}
\begin{split}
p^{r+1}_{z0}- \bar{C}^{r+1}_{z0} 
& = \frac{1}{2}(\hat{C}_{z0}^{r+1}-\bar{C}^{r+1}_{z0})  + \frac{1}{2}( p^r_{0z} + p^r_{1z} - (\bar{C}^r_{0z}+\bar{C}^r_{1z})) \\ & {}- \frac{1}{2} (\hat{C}_{z1}^{r+1} - \bar{C}^{r+1}_{z1} ) + b_z \cdot e_z.
\end{split}
\end{equation}

For indices $i = 1, \dots, r$ and $j = 1, \dots, 2^{i+1}$, let $X_{i, j}$ denote an independent random variable drawn from the discrete Gaussian $\calN(0, \frac{R}{2\rho})$, and let $Y_{i, j} \in \{-1, 1\}$ denote an independent Rademacher random variable.

We claim that the error $p_s^r - \overline{C}_s^r$, denoted by $\Theta^r$, is distributed as
\begin{equation}\label{eq.noise1}
 \Theta^r= \begin{cases}  
 X_{1, 1} & r=0 \\
 \begin{split}
 &\sum_{i = 1}^{r-1} \left(\sum_{j = 1}^{2^i} 2^{-i} X_{i, j} + \sum_{j = 1}^{2^{i-1}} 2^{-i}Y_{i, j} E_{i, j}\right) \\
 &+ \sum_{j = 1}^{2^{r+1}} 2^{-r} X_{r, j} + \sum_{j=1}^{2^{r-1}} 2^{-r}Y_{r, j} E_{r, j}& r\ge 1,
 \end{split}
 \end{cases}
\end{equation}
for some random variables $E_{i, j} \in \{0, 1\}$ that may be arbitrarily correlated with each other, with the $X_{i, j}$'s, and with the $Y_{i', j'}$ for $i' > i$, but are independent from $Y_{i', j'}$ for $i' \le i$.

We will prove Equation \eqref{eq.noise1} by induction. It is easy to verify that when $r=0$, we have $\Theta^0=X_{1, 1}$, 
and when $r=1$, we have from Equation~\eqref{eqn:error-decomp} that $\Theta^1=\sum_{j = 1}^4 \frac{1}{2}X_{1, j} + \frac{1}{2}Y_{1, 1} E_{1, 1}$. 
Suppose Equation \eqref{eq.noise1} holds for update step $r$.
Because of \eqref{eqn:error-decomp} and symmetry of the discrete Gaussian, we have
\begin{align*}
\Theta^{r+1} 
&= \frac{1}{2}X_{1, 1} + \frac{1}{2}(\Theta^r_1+\Theta^r_2)+ \frac{1}{2} X_{1, 2} + \frac{1}{2}Y_{1, 1} E_{1, 1}\\
&=  \sum_{i = 1}^{r} \left(\sum_{j = 1}^{2^i} 2^{-i} X_{i, j} + \sum_{j = 1}^{2^{i-1}} 2^{-i}Y_{i, j} E_{i, j}\right)\\
&+ \sum_{j = 1}^{2^{r+2}} 2^{-(r+1)} X_{r, j} + \sum_{j=1}^{2^{r}} 2^{-(r+1)} Y_{r, j} E_{r, j},
\end{align*}
where $\Theta^r_1, \Theta^r_2$ are independent and distributed as in~\eqref{eq.noise1}. Note that $E_{1, 1}$ may depend arbitrarily on $E_{i, j}, X_{i, j}, Y_{i, j}$ for $i > 1$, but is independent from $Y_{1, 1}$. Hence, Equation \eqref{eq.noise1} holds for update step $r+1$, completing the inductive proof of the claim.

We now analyze the variance of $\Theta^r$ by analyzing the contributions of the discrete Gaussian terms $X_{i, j}$ and the rounding terms $Y_{i, j} E_{i, j}$ separately. Letting
\[G^r =  \sum_{i = 1}^{r-1}\sum_{j = 1}^{2^i} 2^{-i} X_{i, j}  + \sum_{j = 1}^{2^{r+1}} 2^{-r} X_{r, j}\]
and using the fact that each $X_{i, j}$ is an independent discrete Gaussian with variance $\sigma^2=\frac{R}{2\rho}$, we get that the variance of $G^r$ is 
\[\mathrm{Var}(G^r) = \sum_{i = 1}^{r-1}\sum_{j = 1}^{2^i} 2^{-2i} \sigma^2  + \sum_{j = 1}^{2^{r+1}} 2^{-2r} \sigma^2 = \left(2^{-r} + \sum_{i=1}^r 2^{-i}\right) \sigma^2
= \sigma^2,\]
which is the same regardless of the update step $r$.

Denoting the contribution of the rounding terms by
\[B^r = \sum_{i=1}^r \sum_{j=1}^{2^{i-1}} 2^{-i} Y_{i, j} E_{i, j},\]
we observe that $-1/2 \le B^r \le 1/2$ with probability $1$. Moreover, by linearity of expectation and the fact that every $Y_{i, j}$ is independent from $E_{i, j}$, we have that $\mathbb{E}[B^r] = 0$. Therefore, $B^r$ is a subgaussian random variable with variance at most $1/4$, i.e., $\Pr[B^r \ge \lambda] \le e^{-8\lambda^2}$ for all $\lambda \ge 0$.

We now combine the two sources of errors. Since $G^r$ and $B^r$ are (dependent) subgaussians with variances $\sigma^2 = R/2\rho$ and $1/4$ respectively, their summation is subgaussian with variance $(\sigma + 1/2)^2$~\cite{Rivasplata12}.  Thus, we have the following tail bound for every $\lambda \ge 0$.
\begin{equation*}
    \Pr(\abs{p^r_{s}-(C^{r}_{s}+n_{\text{pad}})} \ge \lambda)\le \exp\left(-\frac{\lambda^2 }{2(\sigma + 1/2)^2}\right).
\end{equation*}
Applying a union bound over the total of $2^kR$ bins, we have  
\[
\Pr\left(\bigvee_{\substack{r=0,1,\ldots,R-1\\ s\in\{0,1\}^{k}}} \abs{p^r_s-(C^r_s+n_{\text{pad}})} \ge \lambda \right) \le 2^k R \exp\left(-\frac{\lambda^2 }{2(\sigma + 1/2)^2}\right).
\]
To ensure $2^kR\exp\left(-\frac{\lambda^2}{2(\sigma^2+1/2)}\right) \le \beta$, it suffices to take \\ $\lambda=\left(\sqrt{\frac{R}{\rho}}+\frac{1}{\sqrt{2}}\right)\sqrt{\log(\frac{2^kR}{\beta})}.$ 
Substituting $T-k+1$ back in for $R$ gives: $\lambda=\left(\sqrt{\frac{T-k+1}{\rho}}+\frac{1}{\sqrt{2}}\right)\sqrt{\log(\frac{2^k(T-k+1)}{\beta})}.$
\end{proof}

\begin{corollary} \label{cor:rel-error}
Let $\beta \in (0, 1)$ and \\ $n_{\mathrm{pad}} = \left(\sqrt{\frac{T-k+1}{\rho}}+\frac{1}{\sqrt{2}}\right)\sqrt{\log(\frac{2^k(T-k+1)}{\beta})}$. With probability at least $1-\beta$, the maximum relative error (with respect to the original fraction $C_s^t/n$) is bounded by
\end{corollary}

\begin{equation*}
   \max_{s, t} \left|\frac{p^t_{s}}{n^*}-\frac{C^t_{s}}{n} \right| \le  O\left(\frac{\sqrt{T( k + \log(T / \beta))}}{n \sqrt{\rho}}(1+\frac{2^kC^t_{s}}{n} )\right). 
\end{equation*}

\begin{proof}
For each $s$ and $t$ we may write
\begin{align*}
\left|\frac{p^t_{s}}{n^*}-\frac{C^t_{s}}{n} \right| &\le \left|\frac{p^t_{s}}{n^*}-\frac{(C^t_{s} + n_{\mathrm{pad}})}{n^*} \right| + \frac{n_{\mathrm{pad}}}{n^*} + \left|\frac{C^t_{s}}{n} - \frac{C^t_{s}}{n^*} \right|
\end{align*}
Let $\lambda = \left(\sqrt{\frac{T-k+1}{\rho}}+\frac{1}{\sqrt{2}}\right)\sqrt{\log(\frac{2^k(T-k+1)}{\beta})}$. Condition on the good event that $|p_s^t - (C_s^t + n_{\mathrm{pad}})| \le \lambda$ for every $s$ and $t$, which from the proof of Theorem~\ref{thm.acc2}, happens with probability at least $1-\beta$. Then in particular, we have $n \le n^* \le n + 2^{k+1}\lambda$, which implies that for every $s, t$, this expression is at most
\[\frac{\lambda}{n^*} + \frac{n_{\textrm{pad}}}{n^*} + \frac{C^t_{s}}{n}\frac{n^* - n}{n^*} \le \frac{2\lambda}{n} + \frac{2^{k+1}\lambda}{n} \frac{C^t_{s}}{n}.\]

\end{proof}

This statement gives an upper bound on the worst-case error of the synthetic dataset relative to the original dataset. This $2^k$ dependence affects multiplicative error, so its its main impact is on large histogram bins (with larger $C_s^t$), where the signal is large anyway. On true bins which are “almost average” in the sense that they contain $O(n / 2^k)$ entries, the total additive error is only $\tilde{O}(T\sqrt{k} / n)$.

This answer is biased due to the padding. However, note  that the parameters $n_{\text{pad}}$ and $k$ are public. So an analyst with knowledge of these parameters can debias each query answer by subtracting $n_{\text{pad}}$ from each noisy count. The resulting maximum relative error is bounded by $  \max_{s, t} \frac{\abs{(p^t_{s}-n_{\mathrm{pad}})-C^{t}_{s}}}{n}\le \dfrac{\left(\sqrt{\frac{T-k+1}{\rho}}+\frac{1}{\sqrt{2}}\right)\sqrt{\log\left(\frac{2^k(T-k+1)}{\beta}\right)}}{n}. $

An interesting feature of our accuracy analysis is that it makes some use of the distribution of errors introduced in the noisy histogram computation, rather than just a black-box upper bound on the magnitude of the error. Indeed, one could analyze accuracy by applying the triangle inequality to the magnitudes of these errors, but it would lead to worse error bounds.

\section{Generating Synthetic Data for Cumulative Time Queries}

In this section, we consider cumulative time queries, which track the Hamming weight of each individual's history. Specifically, we would like the synthetic data to approximately preserve the fraction of people having at least $b$ 1's in their data up to time $t$, i.e., $c_b^t(D^1, \dots, D^t) = \frac{1}{n}\#\{i| x_i^1 + x_i^2 +\ldots+ x_i^t \ge b\}$ for $t=1, 2, \ldots, T$. 
In Section 2.1, we reduced cumulative time queries to fixed-time window queries. The worst-case error is $\widetilde{O}(T / n)$ when instantiating the algorithm for fixed-time window queries with $k=T$. Here, we provide a CDP algorithm for this problem with a better accuracy guarantee of $\widetilde{O}(\sqrt{T} / n)$. 

\forappendix{
\subsection{Preliminaries on Stream Counters}

Our algorithm for cumulative time queries relies on the following important primitive from the literature on private continual release. A \emph{stream counter} takes as input a stream of values $z^1, \dots, z^T$, where each $z^t$ is a natural number. At every time step $t$, it releases an approximation $\hat{S}^t$ to the partial sum $S^t = \sum_{k = 1}^t z^t$. Two streams $(z^1, \dots, z^T), ((z^1)', \dots, (z^T)')$ are neighbors if they differ in only one entry, and by magnitude $1$ in that entry. That is, there exists an index $t$ such that $|z^t - (z^t)'| \le 1$ and $z^k = (z^k)'$ for all $k \ne t$. We are interested in stream counters whose sequence of outputs $(\widetilde{S}^1, \dots, \widetilde{S}^T)$ guarantee $\rho$-zCDP with respect to this neighboring relation. Moreover, we define accuracy for stream counters as follows. 

\begin{definition}
    A stream counter is $(\alpha(\rho,T), \beta(\rho, T))$-accurate if for every stream of length $T$ and every $\rho$, it holds that
    \begin{equation*}
        \Pr[ \abs{\widetilde{S}^t-S^t} \le \alpha(\rho,T) ]\ge 1-\beta(\rho, T),
    \end{equation*}
for every $1\le t\le T$.
\end{definition}

Note that in contrast to our definition of $(\alpha, \beta)$-accuracy for synthetic data generation, here $\alpha(\rho, T)$ is generally in the range $[0, T]$, i.e., it represents error with respect to counts rather than with respect to fractions.

The {\em tree-based aggregation mechanism}~\cite{DNPR10,chan2011private} was the first implementation of a private stream counter.
The algorithm works roughly as follows. Consider a complete binary tree, in which each leaf node is labeled by $z^t$ and represents the data arriving single time step. Each internal node tracks the sum of all leaves in its sub-tree. To ensure privacy, we add fresh discrete Gaussian noise with scale $\sigma^2=\log T/(2\rho)$, denoted as $\calN(0,\log T/(2\rho)) $ to each internal node. Because every time step only affects $O(\log T)$ nodes in this binary tree, the complete tree is $\rho$-zCDP by composition. 

\begin{theorem}
\label{thm.privtree}

The tree-based aggregation mechanism is $\rho$-zCDP.
\end{theorem}

Every partial sum $S^t$ can be expressed as a sum of $O(\log t)$ nodes in this tree, and each such node introduces fresh noise $\mathcal{N}_{\mathbb{Z}}(0,\log T/(2\rho))$. Therefore, the error for computing a private version of the partial sum $\abs{\widetilde{S}^t-S^t}$ is $O(\sqrt{\frac{\log T}{\rho}} \sqrt{\log t})$ with high probability. We formally describe the algorithm in Algorithm \ref{algo:tree}. (Note that the tree-based aggregation algorithm was initially described using Laplace noise, resulting a pure $(\eps, 0)$-DP algorithm \cite{DNPR10,chan2011private}.)

\begin{theorem}
\label{thm.acctree}
For each $t=1, \ldots, T$ individually, with probability at least $1-\beta$, the error of the tree-based aggregation mechanism is bounded as follows:
\begin{equation*}
    \abs{\widetilde{S}^t-S^t} \le O\left(\sqrt{\frac{\log T}{\rho}}\cdot \sqrt{\log t} \cdot \log \frac{1}{\beta}\right).
\end{equation*}
\end{theorem}

\begin{algorithm}[ht]
\caption{Tree-based aggregation }\label{algo:tree}
\begin{algorithmic} 
\State \textbf{Input:} Time upper bound $T$, privacy parameter $\rho$, and a stream $z=(z^1, z^2, \ldots, z^T)\in \mathbb{N}^T$.
\State \textbf{Output:} a noisy partial sum $\widetilde{S}^t$ at each time step $t$.
\State \textbf{Initialization:} Each $\alpha_i$ and $\widetilde{\alpha}_i$ are initialized to $0$.

\For{ $t\gets 1$ to $T$ } 
 \State Express $t$ in binary form: $t=\sum_j \textbf{Bin}_j(t) \cdot 2^j$.

 \State Let $i = \min\{j: \textbf{Bin}_j(t) \neq 0\}$.

 \State Set $\alpha_i \gets \sum_{j<i}\alpha_j + z^t$.

 \For{ $j\gets 0$ to $i-1$}
  \State Update $\alpha_j\gets 0$ and $\hat{\alpha}_j \gets 0$.
 \EndFor

 \State Update $\widetilde{\alpha}_i\gets \alpha_i+\calN(0,\frac{\log T}{2\rho})$.

 \State Calculate the noisy partial sum at time $t$: $\widetilde{S}^t\gets \sum_{j:\textbf{Bin}_j(t)=1} \widetilde{\alpha}_j$.

\EndFor

\end{algorithmic} 
  \end{algorithm}
}

\subsection{Algorithm}

Our algorithm, presented formally in Algorithm \ref{algo:cumulative}, 
 again consists of two phases: producing private statistics about the cumulative queries and constructing synthetic data according to the noisy statistics.

\begin{algorithm}[ht]
\caption{Private synthetic data generation preserving cumulative queries.} \label{algo:cumulative}
\begin{algorithmic}

\State  \textbf{Input:} Dataset consisting of $n$ users/streams $(D^1, D^2, \ldots, D^T)$ where $D^t=(x_1^t, \ldots, x_n^t)$ is the $t$-th vector of data produced at time $t$,
  time upper bound $T$, stream counter mechanism $\mathcal{M}$ with a sequence of privacy parameters $(\rho_1, \rho_2, \ldots, \rho_T)$ such that $\sum_{b=1}^T \rho_b=\rho$.
\State  \textbf{Output:} A vector $\hat{D}^t=(\hat{x}_1^t, \ldots, \hat{x}_n^t)$ at each time step $t$.
\State  


\For{ $t\gets 1$ to $T$ } 

\State Initialize $\hat{D}^t=(0, 0, \ldots, 0)^T$

\For{$b\gets 0$ to $t$} 

\State Set $z_b^t= \#\{i\mid\{x_i^1 + \ldots + x_i^{t-1} = b-1 \text{ and } x^{t}_i=1\}$.

\State Invoke the stream counter $\mathcal{M}_b$ to calculate the noisy \State \hspace{2mm} counts at time $t$:  $\tilde{S}_b^t\gets \mathcal{M}_b(z_b^1, \dots, z_b^t) $.

\State Set $\hat{S}_b^t = \min\{\max\{\tilde{S}_b^t, \hat{S}_b^{t-1}\}, \hat{S}_{b-1}^{t-1}\}.$


\State Calculate $\hat{z}_b^t=\hat{S}_b^t-\hat{S}_b^{t-1}$.

\State Randomly select $\hat{z}_b^t$ indices, denoted $I_b$, from the index \State \hspace{2mm} set 
 $I = \{i\mid x_i^1 + \ldots + x_i^{t-1} = b-1 \}$.

\State Update $\hat{x}_i^t\gets 1$ for $i \in I_b$ and $\hat{x}_i^t \gets 0$ for $i \in I \setminus I_b$.

\EndFor

\EndFor

\end{algorithmic}
  \end{algorithm}

\paragraph{Producing noisy statistics.}

Our algorithm relies on a stream counter mechanism $\mathcal{M}$, which takes in a stream of variables $z^1, z^2, \ldots, z^T$ and produces a private estimate of the sum $S^t=\sum_{j=1}^t z^j$ at each time $t$. For each $b = 0, 1, \dots, T$, we create an instance $\mathcal{M}_b$ of the stream counter to track the number of people having at least $b$ 1's up to time $t$, denoted by $S_b^t=\#\{i| x_i^1 + x_i^2 +\ldots+ x_i^t \ge b\}$ for each $t=1, 2, \ldots, T$. To see how to do this, let $z_b^t= \#\{i|\{x_i^1 + \ldots + x_i^{t-1} = b-1 \text{ and } x^{t}_i=1\}$, which denotes the number of people with exactly $b-1$ 1's up to time $t-1$, and also have a $1$ at time $t$. This allows us to represent $S_b^t$ as the sum $S_b^t=\sum_{j=1}^t z_b^j$. Note that for each value of $b$, every user contributes at most one value of $z_b^t$. This ensures that neighboring datasets induce neighboring streams for each stream counter. In Section \ref{sec.exp}, we instantiate the stream counter using the tree-based aggregation mechanism in our experiments.

The true counts $(S_b^{1}, \dots, S_b^{T})$ are monotonically increasing because each $z_b^t$ is nonnegative. But adding noise for privacy may violate the monotonicity of the stream. Maintaining monotonicity is essential for ensuring the existence of consistent synthetic data in the next phase. To ensure monotonicity, we maintain a monotonized stream as follows. In each time step $t$, let $\tilde{S}_b^t$ be the output of the stream counter $\mathcal{M}_b$. Set the output $\hat{S}_b^t$ of the monotonized counter to be $\hat{S}_b^t = \min\{\max\{\tilde{S}_b^t, \hat{S}_b^{t-1}\}, \hat{S}_{b-1}^{t-1}\}$, i.e., clamp it so that $\hat{S}_b^{t-1} \le \hat{S}_b^t \le \hat{S}_{b-1}^{t-1}$. We require the upper bound $\hat{S}_b^t \le \hat{S}_{b-1}^{t-1}$  because the number of people having $b$ 1's up to time $t$ cannot exceed the number of people having $b-1$ 1's up to time $t-1$. A similar idea for maintaining consistency for a single stream counter was shown in \cite{chan2011private} not to increase the error in any of the counts produced. Our generalization of this fact shows that it continues to hold even when we require a natural monotonicity constraint across \emph{different} invocations of the stream counter.

Below, we show that our monotonization procedure does not increase the worst-case ($\ell_\infty$) error of the noisy counts. In~\cite{HayRMS10}, it was shown how to use a different monotonization procedure to preserve the $\ell_2$ error of the vector of noisy counts. However, their procedure appears to make essential use of the fact that all of the data is available at once, and it is an interesting question to determine whether one can obtain similar guarantees using an algorithm that operates on data arriving incrementally.

\paragraph{Generating synthetic data.} By post-processing, we may also privately reveal the statistics $\hat{z}_b^t=\hat{S}_b^t-\hat{S}_b^{t-1}$. At time $t$, we update the synthetic dataset to maintain consistency with $\hat{z}_b^t$ for every $b=1,\ldots, t$. To be concrete, at time $t$, for the set of indices $\{i\mid \hat{x}_i^1 + \ldots + \hat{x}_i^{t-1} = b-1\}$, the algorithm extends $\hat{z}_b^t$ rows by $1$, and the remaining $\hat{S}_{b-1}^{t-1} - \hat{S}_b^t$ rows by 0. This is possible because the monotonization procedure described above ensures that $\hat{z}_b^t \ge 0$ and $ \hat{S}_{b-1}^{t-1} - \hat{S}_b^t \ge 0$. 

\subsection{Theoretical guarantees}

\paragraph{Privacy.}
The privacy of Algorithm \ref{algo:cumulative} follows from the composition of $T$ stream counters.

\begin{theorem}
Algorithm \ref{algo:cumulative} satisfies $\rho$-zCDP.
\end{theorem}

\paragraph{Accuracy.} Our algorithm monotonizes the stream of counters to be $\hat{S}_b^t = \min\{\max\{\tilde{S}_b^t, \hat{S}_b^{t-1}\}, \hat{S}_{b-1}^{t-1}\}$. Our proof relies on Lemma \ref{lemma1}, showing that the error of the monotonized counter is at most that of the ordinary stream counter, so this step will not incur additional error. 

\begin{lemma}\label{lemma1}
Let the true count at time $t$ be $S_b^t=\#\{i| x_i^1 + x_i^2 +\ldots+ x_i^t \ge b\}$, the noisy count be $\tilde{S}_b^t$, and the monotonized noisy count be $\hat{S}_b^t$.
For every $t=1,2,\ldots, T$, and every $b=1, 2, \ldots, t$, we have 
\begin{equation}\label{lemma.error}
\abs{\hat{S}_b^t-S_b^t}\le \max \{ \abs{\tilde{S}_b^t-S_b^t}, \abs{\hat{S}_b^{t-1}-S_b^{t-1}}, \abs{\hat{S}_{b-1}^{t-1}-S_{b-1}^{t-1} }  \}. 
\end{equation}
For $b=0$, 
\begin{equation}\label{lemma.error0}
\abs{\hat{S}_b^t-S_b^t}\le \max \{ \abs{\tilde{S}_b^t-S_b^t}, \abs{\hat{S}_b^{t-1}-S_b^{t-1}} \}. 
\end{equation}
\end{lemma}

\begin{proof}
The proof of Lemma \ref{lemma1} relies on the following simple inequality.
\begin{claim}\label{helper.claim}
For real numbers $A, B, C, D$, if $A\le B$, and $D\le C$, then 
\begin{equation*}
 \abs{B-C}\le \max \{ \abs{A-C}, \abs{B-D}\}.  
\end{equation*}
\end{claim}

We first prove this claim. Fix $A$ and $B$. Depending on $C$, we consider the following three cases. When $A \le B \le C$, $\abs{B-C}\le \abs{A-C}$. When $C \le  A \le B$, $\abs{B-C}\le \abs{B-D}$. When $A \le C \le B$, $\abs{B-C}\le \abs{B-D}$ holds regardless of where $D$ is relative to $A$, so we complete the proof for this claim.

Lemma \ref{lemma1} proceeds from Claim \ref{helper.claim}, which can be seen as follows. When $\hat{S}_b^t=\tilde{S}_b^t$, then accuracy follows from that of the ordinary stream counter. Thus we only need to consider the two cases where $\hat{S}_b^t=\hat{S}_b^{t-1}$ or $\hat{S}_b^t=\hat{S}_{b-1}^{t-1}$. For the first case where $\hat{S}_b^t=\hat{S}_b^{t-1}$, 
we have $\tilde{S}_b^t \le \hat{S}_b^t$ by the construction of $\hat{S}_b^t$, and $S_b^{t-1}\le S_b^t$ because of the monotonicity of $S_b^t$. Instantiating Claim \ref{helper.claim} with $A=\tilde{S}_b^t$, $B=\hat{S}_b^t$, $C=S_b^t$, and $D=S_b^{t-1}$, we have 
\begin{equation}\label{eq.1}
\begin{split}
\abs{\hat{S}_b^t-S_b^t} &\le \max \{ \abs{\tilde{S}_b^t-S_b^t}, \abs{\hat{S}_b^{t}-S_b^{t-1}} \\&
= \max \{ \abs{\tilde{S}_b^t-S_b^t}, \abs{\hat{S}_b^{t-1}-S_b^{t-1}} \}. 
\end{split}
\end{equation}
Similarly, for the second case where $\hat{S}_b^t=\hat{S}_{b-1}^{t-1}$, we have $\tilde{S}_b^t \ge \hat{S}_b^t=\hat{S}_{b-1}^{t-1}$ by the construction of $\hat{S}_b^t$, and $S_{b-1}^{t-1}\ge S_b^t$ because of the monotonicity of $S_b^t$.
We then instantiate Claim \ref{helper.claim} with $A=-\tilde{S}_b^t$, 
$B=-\hat{S}_{b-1}^{t-1}$, $C=-S_b^t$ and $D=-S_{b-1}^{t-1}$. 
and we have 
\begin{equation}\label{eq.2}
\abs{\hat{S}_{b-1}^{t-1}-S_b^t}\le \max \{ \abs{\tilde{S}_b^t-S_b^t}, \abs{\hat{S}_{b-1}^{t-1}-S_{b-1}^{t-1} }  \}. 
\end{equation}
Since $\hat{S}_b^t=\hat{S}_{b-1}^{t-1}$, ~\eqref{eq.2} is equivalent to 
\begin{equation}\label{eq.3}
\abs{\hat{S}_{b}^{t}-S_b^t}\le \max \{ \abs{\tilde{S}_b^t-S_b^t}, \abs{\hat{S}_{b-1}^{t-1}-S_{b-1}^{t-1} }  \}. 
\end{equation}

Combining \eqref{eq.1} and \eqref{eq.3}, we obtain \eqref{lemma.error}, completing the proof.

\end{proof}

Lemma \ref{lemma1} implies that the worst-case error of Algorithm \ref{algo:cumulative} is at most the worst-case error of $T$ stream counters, normalized by the dataset size $n$.

\begin{theorem}\label{thm.cumulative-error}
When instantiated with an $(\alpha, \beta)$-accurate stream counter $\mathcal{M}$, Algorithm \ref{algo:cumulative} is $(\alpha^*, \beta^*)$-accurate
for\\ $\alpha^*= \frac{1}{n}\max_b \alpha\left(\rho_b, T-b+1\right)$ and
    $\beta^*= \sum_{b=1}^T \beta(\rho_b, T-b+1).$
\end{theorem}

\forshort{In the Appendix, we state the accuracy guarantee when we instantiate the tree-based aggregation mechanism as the stream counter $\mathcal{M}$.}

\section{Illustrating Examples}\label{sec.exp}

In this section, we provide examples to illustrate our approaches.
We apply our algorithms to data from the Survey of Income and Program Participation (SIPP) by the U.S. Census Bureau~\cite{sipp2021}. Unless otherwise stated, we use noise-adding mechanisms with the variance $\sigma^2$ calibrated to achieve overall $0.005$-zCDP. \forshort{Further examples with simulated data and varying privacy parameters are included in the Appendix.} Code to replicate the experiments is part of the Supporting Information of this paper.

\paragraph{Real Data Application: Releasing Poverty Rates from the Survey of Income and Program Participation (SIPP)}
We apply our Algorithm \ref{algo:fixedwindow} to data collected in the SIPP~\cite{sipp2021}. We downloaded the 2021 sample from \url{https://www2.census.gov/programs-surveys/sipp/data/datasets/2021/pu2021_csv.zip} and detail the pre-processing steps below. We aim to estimate the proportion of households in the SIPP data that are in poverty.

We need to take a couple of data pre-processing steps to get the SIPP data into shape for our data synthesizer. First, in the SIPP data, it is possible that multiple persons per household are surveyed. Thus, we first subset the data to one longitudinal series per household. The SIPP variable `THINCPOVT2' is coded as the household income ratio to the household  poverty threshold in a given month. We binarize this such that any values of the ratio smaller than one are coded as $1$ (indicating that a household was in poverty in a given month). Finally, some households have missing values. We delete every household that has at least one missing value.

Our final SIPP data sample consists of $23374$ households ($N = 23374$) with  $12$ monthly measurements ($T = 12$) for 2021 that indicate whether a household was in poverty in a given month. For our experiment, we treat this SIPP sample as ground truth and we focus on one particular poverty indicator, ``Household income-to-poverty ratio in this month, including Type 2 individuals''.

\paragraph{Synthetic data for fixed time window queries.} To answer fixed time window queries, we synthesize the SIPP sample with a window size of 3 ($k = 3$), capturing  quarterly trends. We  repeat the synthesizer $1000$ times. To estimate the proportion of households in poverty in a given quarter, it is natural to examine four different quantities. First, the proportion of households in poverty for at least one month of the quarter. Second, the proportion of households in poverty for at least two months of the quarter. Third, the proportion of households in poverty for at least two consecutive months of the quarter. And fourth, the proportion of households in poverty for all three months of the quarter. \forshort{We present the answers based on the synthetic data against the ground truth in Figure \ref{fig:figure1}.}\forappendix{We present the results in Figures \ref{fig-app:experiment31}-\ref{fig-app:experiment33} for different values of the privacy parameter $\rho$.} \forappendix{In the right panels of Figures \ref{fig-app:experiment32}-\ref{fig-app:experiment33}, we show that our answers based on the synthetic data averaged over 1000 repetitions accurately match the ground truth, indicating that our approach provides unbiased estimate of the fixed time queries (by subtracting the query answer on the padding data from the query answer on the complete synthetic data).} 

\forshort{This shows that the synthetic data has some utility. However, }\forappendix  {In the left panels of Figures \ref{fig-app:experiment32}-\ref{fig-app:experiment33}, we show that} ignoring the debiasing step can lead to substantial bias. \forshort{As the amount of padding is public information, applying the debiasing step is easy and lets analysts recover unbiased results. In the Appendix, we show more results calculated directly on the synthetic data and their debiased counterparts.}
A major advantage of our synthetic data approach is that we can answer any such query written as a linear combination of histogram queries (with a fixed window size of at most $k$) without any additional privacy cost. 

\forshort{

\begin{figure}[ht]
\vskip 0.2in
\begin{center}
\centerline{\includegraphics[width=0.7\linewidth]{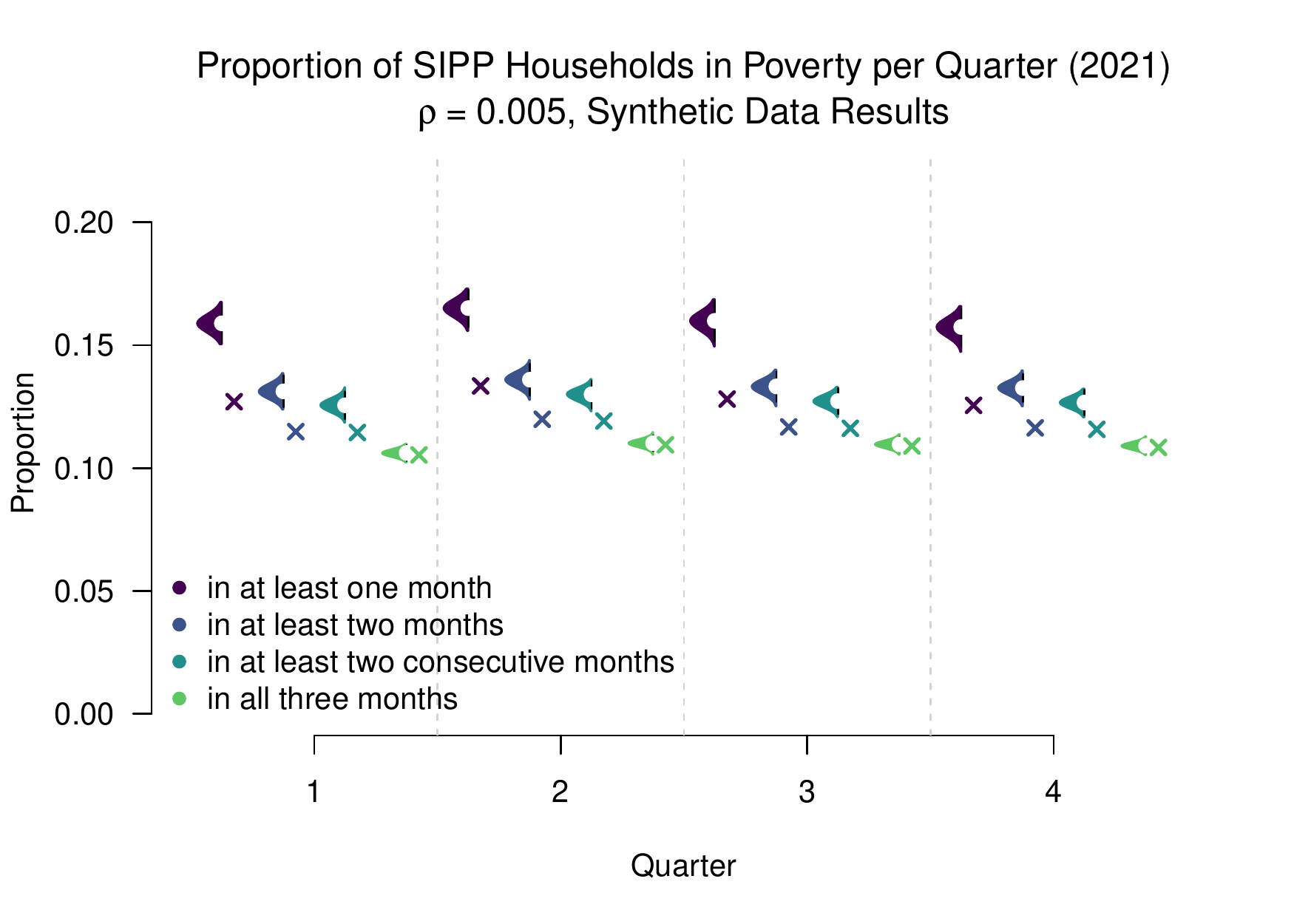}}
\caption{Proportions of SIPP Households in poverty per quarter in 2021. Calculated on the synthetic data. the density estimates show the empirical privacy noise distribution across $1000$ repetitions of the experiments with privacy parameter $\rho = 0.005$. X's indicate the ground truth calculated from the SIPP data.}
\label{fig:figure1}
\end{center}
\vskip -0.2in
\end{figure}
\begin{figure}[ht]
\vskip 0.2in
\begin{center}
\centerline{\includegraphics[width=0.7\linewidth]{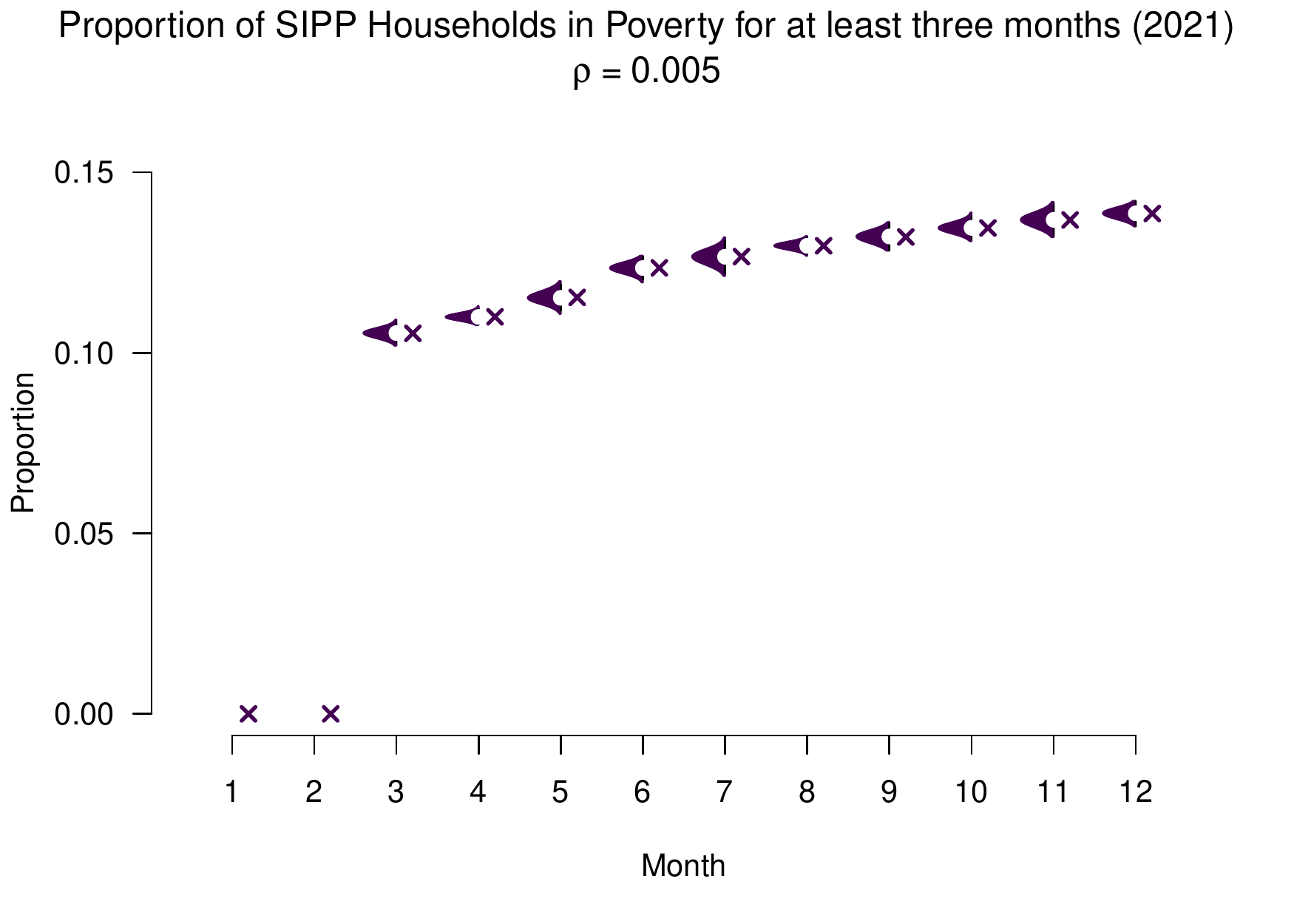}}
\caption{Proportion of SIPP Households in poverty for at least three months up to any given month in 2021. In both panels, the density estimates show the empirical privacy noise distribution across $1000$ repetitions of the experiments with privacy parameter $\rho = 0.005$. X's indicate the ground truth calculated from the SIPP data.}
\label{fig:figure2}
\end{center}
\vskip -0.2in
\end{figure}
}

\paragraph{Synthetic data based on cumulative time queries.}
Finally, we apply Algorithm \ref{algo:cumulative} to synthesize the SIPP sample. With the DP synthetic data based on cumulative time queries, we can answer queries like ``How many households were at least three months in poverty in any given month''. \forshort{In Figure \ref{fig:figure2}}\forappendix{In Figure \ref{fig-app:experiment4}}, we again repeat the synthesizer 1000 times and plot the answers to these queries using our generated DP synthetic data versus the ground truth, with the x-axis denoting the time horizon. Our answers based on the synthetic data averaged over 1000 repetitions accurately match the ground truth, indicating that our approach provides an unbiased estimate of the cumulative time queries.
\forappendix{While Algorithm \ref{algo:cumulative} generates synthetic data for all time thresholds $b$ from $1, \dots, T$ simultaneously, we here focus on the results for setting the threshold to $b = 3$.}

\section{Conclusion}
In this work we have studied the problem of continuously
releasing differentially private synthetic data from longitudinal studies. We have defined a formal model for this problem and showed two instances for two class of queries inspired by census-like longitudinal data. We have provided theoretical upper bounds for these two classes of queries and showed through examples how they can be used in practice on census data. 

Synthetic data can enable analysts to answer supported queries without prior knowledge and without modifying their analysis pipelines. This is the main reason behind the demand for tabular synthetic data and why it is a legal requirement for certain data releases. Nevertheless, synthetic data is intrinsically limited in the scope of analyses it permits. It is important to acknowledge that synthetic data may not always match the accuracy of DP query answers that can be released through some other data structure~\cite{Abowd2021}. 
Our work identifies restricted but expressive classes of queries that permit accurate synthetic data generation algorithms. This approach aligns with the existing literature on DP single-shot synthetic data release, which provides theoretical accuracy guarantees for specific query classes. We envision restricted synthetic data as just part of a larger privacy-preserving data access system, where after performing initial exploratory data analysis on synthetic data, analysts may then decide to go through additional approvals to access the original data or to more accurate estimates.

\subsubsection*{Acknowledgments}
We thank Salil Vadhan for helpful comments on the presentation of this paper. The research presented in this paper was supported by the U.S. Census Bureau Cooperative Agreement CB20ADR0160001. W.Z. is supported in part by a Computing Innovation Fellowship from the Computing Research Association (CRA) and the Computing Community Consortium (CCC). M.N. started to work on this project while at Boston University.

\forappendix{
\small

\clearpage

}

\newpage


\printbibliography

\appendix

\section{Preliminaries on Stream Counters}


Our algorithm for cumulative time queries relies on the following important primitive from the literature on private continual release. A \emph{stream counter} takes as input a stream of values $z^1, \dots, z^T$, where each $z^t$ is a natural number. At every time step $t$, it releases an approximation $\hat{S}^t$ to the partial sum $S^t = \sum_{k = 1}^t z^t$. Two streams $(z^1, \dots, z^T), ((z^1)', \dots, (z^T)')$ are neighbors if they differ in only one entry, and by magnitude $1$ in that entry. That is, there exists an index $t$ such that $|z^t - (z^t)'| \le 1$ and $z^k = (z^k)'$ for all $k \ne t$. We are interested in stream counters whose sequence of outputs $(\widetilde{S}^1, \dots, \widetilde{S}^T)$ guarantee $\rho$-zCDP with respect to this neighboring relation. Moreover, we define accuracy for stream counters as follows. 

\begin{definition}
    A stream counter is $(\alpha(\rho,T), \beta(\rho, T))$-accurate if for every stream of length $T$ and every $\rho$, it holds that
    \begin{equation*}
        \Pr[ \abs{\widetilde{S}^t-S^t} \le \alpha(\rho,T) ]\ge 1-\beta(\rho, T),
    \end{equation*}
for every $1\le t\le T$.
\end{definition}

Note that in contrast to our definition of $(\alpha, \beta)$-accuracy for synthetic data generation, here $\alpha(\rho, T)$ is generally in the range $[0, T]$, i.e., it represents error with respect to counts rather than with respect to fractions.

The {\em tree-based aggregation mechanism}~\cite{DNPR10,chan2011private} was the first implementation of a private stream counter. 
The algorithm works roughly as follows. Consider a complete binary tree, in which each leaf node is labeled by $z^t$ and represents the data arriving single time step. Each internal node tracks the sum of all leaves in its sub-tree. To ensure privacy, we add fresh discrete Gaussian noise with scale $\sigma^2=\log T/(2\rho)$, denoted as $\calN(0,\log T/(2\rho)) $ to each internal node. Because every time step only affects $O(\log T)$ nodes in this binary tree, the complete tree is $\rho$-zCDP by composition. 

\begin{theorem}
\label{thm.privtree}

The tree-based aggregation mechanism is $\rho$-zCDP.
\end{theorem}

Every partial sum $S^t$ can be expressed as a sum of $O(\log t)$ nodes in this tree, and each such node introduces fresh noise $\mathcal{N}_{\mathbb{Z}}(0,\log T/(2\rho))$. Therefore, the error for computing a private version of the partial sum $\abs{\widetilde{S}^t-S^t}$ is $O(\sqrt{\frac{\log T}{\rho}} \sqrt{\log t})$ with high probability. We formally describe the algorithm in Algorithm \ref{algo:tree}. (Note that the tree-based aggregation algorithm was initially described using Laplace noise, resulting a pure $(\eps, 0)$-DP algorithm \cite{DNPR10,chan2011private}.)

\begin{theorem}
\label{thm.acctree}
For each $t=1, \ldots, T$ individually, with probability at least $1-\beta$, the error of the tree-based aggregation mechanism is bounded as follows:
\begin{equation*}
    \abs{\widetilde{S}^t-S^t} \le O\left(\sqrt{\frac{\log T}{\rho}}\cdot \sqrt{\log t} \cdot \log \frac{1}{\beta}\right).
\end{equation*}
\end{theorem}

\begin{algorithm}[ht]
\caption{Tree-based aggregation }\label{algo:tree}
\begin{algorithmic} 
\State \textbf{Input:} Time upper bound $T$, privacy parameter $\rho$, and a stream $z=(z^1, z^2, \ldots, z^T)\in \mathbb{N}^T$.
\State \textbf{Output:} a noisy partial sum $\widetilde{S}^t$ at each time step $t$.
\State \textbf{Initialization:} Each $\alpha_i$ and $\widetilde{\alpha}_i$ are initialized to $0$.

\For{ $t\gets 1$ to $T$ } 
 \State Express $t$ in binary form: $t=\sum_j \textbf{Bin}_j(t) \cdot 2^j$.

 \State Let $i = \min\{j: \textbf{Bin}_j(t) \neq 0\}$.

 \State Set $\alpha_i \gets \sum_{j<i}\alpha_j + z^t$.

 \For{ $j\gets 0$ to $i-1$}
  \State Update $\alpha_j\gets 0$ and $\hat{\alpha}_j \gets 0$.
 \EndFor

 \State Update $\widetilde{\alpha}_i\gets \alpha_i+\calN(0,\frac{\log T}{2\rho})$.

 \State Calculate the noisy partial sum at time $t$: $\widetilde{S}^t\gets \sum_{j:\textbf{Bin}_j(t)=1} \widetilde{\alpha}_j$.

\EndFor

\end{algorithmic} 
  \end{algorithm}

\section{Accuracy guarantee for instantiation using the tree-based aggregation mechanism}

We now state the accuracy guarantee when we instantiate the tree-based aggregation mechanism as the stream counter $\mathcal{M}$.
For the $b$-th tree-based counter, we add $\calN
(0, \sigma_b^2)$ to the internal nodes. The partial sum $\hat{S}_b^t$ is
the summation of at most $\max(\lceil \log_2 (t-b+1) \rceil, 1)$ i.i.d. discrete Gaussian random variables with scale $\sigma_b^2$, which is also Gaussian distributed with variance at most $\sigma_b^2\max(\lceil \log_2 (t-b+1) \rceil, 1)$, where $\lceil x \rceil$ is the ceiling function. By Gaussian tail bound, for every $\alpha>0$, we have
\begin{equation*}
\Pr(\abs{\hat{S}_b^t-S_b^t} \ge \alpha) \le \exp\left(-\frac{\alpha^2}{2 \sigma_b^2\max(\lceil \log_2 (t-b+1) \rceil, 1)}\right).
\end{equation*}
By a union bound over all $t = 1, \dots, T$,  with probability at least $1-\beta$, we have that \\$\abs{\hat{S}_b^t-S_b^t} \le \max(\lceil \log_2 (t-b+1) \rceil, 1)\sqrt{\frac{ \max(\lceil \log_2 (T-b+1) \rceil, 1) }{\rho_b}  \log \frac{1}{\beta}}$ for all $t$ simultaneously. We will choose the privacy parameter for each tree-based counter $\rho_b$, for $b=1,2,\ldots, T$,  to equalize the upper bounds of the worst-case errors of all the counters:
\begin{equation*}
    \frac{\rho_1}{ \max(\lceil \log_2 (T-1+1) \rceil, 1)^3} = \frac{\rho_2}{ \max(\lceil \log_2 (T-2+1) \rceil, 1)^3} = \ldots  =\frac{\rho_T}{1}  .
\end{equation*}
Hence, we set $\rho_b = \rho \frac{\max(\lceil\log_2(T-b+1)\rceil, 1)^3 }{\sum_{b=1}^T \max(\lceil\log_2(T-b+1)\rceil, 1)^3}$. 
\begin{corollary}\label{cor:D7}
When instantiated with tree-based stream counters with $\rho_b =\rho \frac{\max(\lceil\log_2(T-b+1)\rceil, 1)^3 }{\sum_{b=1}^T \max(\lceil\log_2(T-b+1)\rceil, 1)^3}$, $b=1, 2,\ldots, T$, for every $\beta \in (0, 1)$, Algorithm \ref{algo:cumulative} is $(\alpha^*, \beta^*)$-accurate
for 
\begin{align*}
    \alpha^*&= \frac{1}{n} \sqrt{ \frac{\sum_{b=1}^T \max(\lceil\log_2(T-b+1)\rceil, 1)^3}{\rho} \log \frac{1}{\beta}}    \\
    \beta^*&= T\beta.
\end{align*}
\end{corollary}

\section{Further Illustrating Examples}
\subsection{Simulated Data}
We evaluate Algorithm \ref{algo:fixedwindow} on rather extreme simulated data. We set the number of observations $n$ to $25000$   and the time horizon $T$ to $12$. We choose these values to demonstrate typical data sizes in longitudinal studies (e.g., $25000$ survey respondents with monthly measurements for one year). In the extreme setting, all data updates are set to $1$. We then generate synthetic  data to preserve fixed-time window queries using a window size $k$ of $3$ (e.g., to reflect an interest in quarterly statistics). 

\begin{figure}[ht]
\vskip 0.2in
\begin{center}
\centerline{\includegraphics[width=0.5\columnwidth]{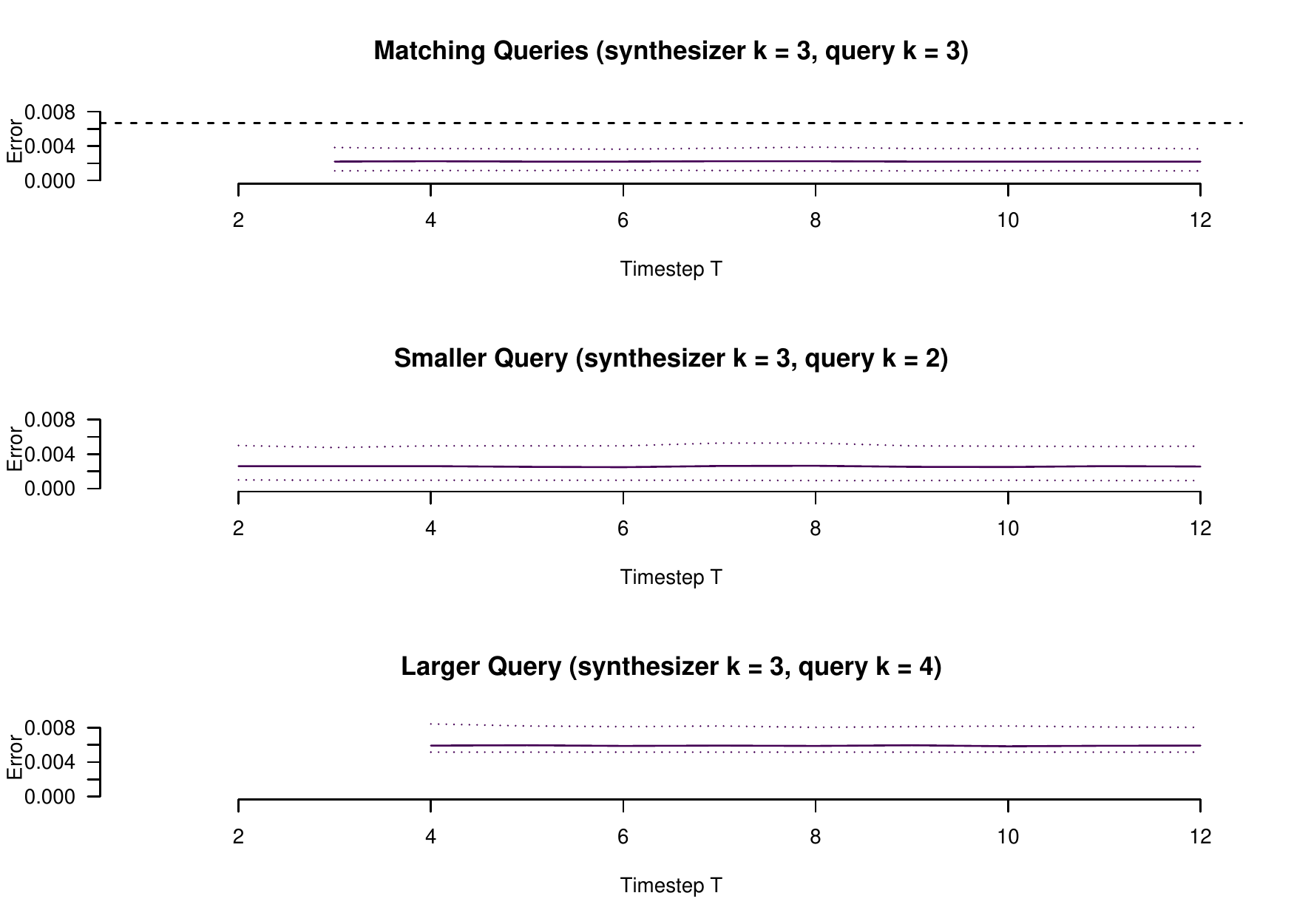}}
\caption{Empirical evaluation of the error of Algorithm \ref{algo:fixedwindow} on simulated data. The solid line shows the median error at each timestep across $1000$ repetitions of the algorithm. The proportions are calculated with the debiasing step. The dotted lines show the $2.5$ and $97.5$ percentile. The horizontal dashed line shows the theoretical error bound. }
\label{fig:experiment1}
\end{center}
\vskip -0.2in
\end{figure}

Figure \ref{fig:experiment1} summarizes the results of $1000$ repeated runs of Algorithm \ref{algo:fixedwindow}. In the top panel of Figure \ref{fig:experiment1}, we show the (desired) scenario where the fixed window size $k$ specified to the synthesizer coincides with the window size of the query we are evaluating. As predicted by Theorem \ref{thm.acc2}, the empirical error remains constant across time. In the middle panel, we highlight an advantage of releasing synthetic and padding data. Since all fixed time window queries with a window size at most $k$ of the synthesizer can be written as a low-weight linear combination of width-$k$ queries, the synthesized data remains accurate for these queries as well. Finally, we display what happens if the window size of the query exceeds $k$ of the synthesizer. Given our algorithm, we do not expect the synthetic data to preserve queries of longer window sizes. Indeed, we can observe that the empirical error increases substantially. We suggest that this warrants a word of caution for synthetic data analysts: Only queries supported by the synthesizer can be answered accurately. We cannot expect reasonable accuracy guarantees for queries not specifically supported by the synthesizer. 

\begin{figure}[ht]
\vskip 0.2in
\begin{center}
\centerline{\includegraphics[width=0.5\columnwidth]{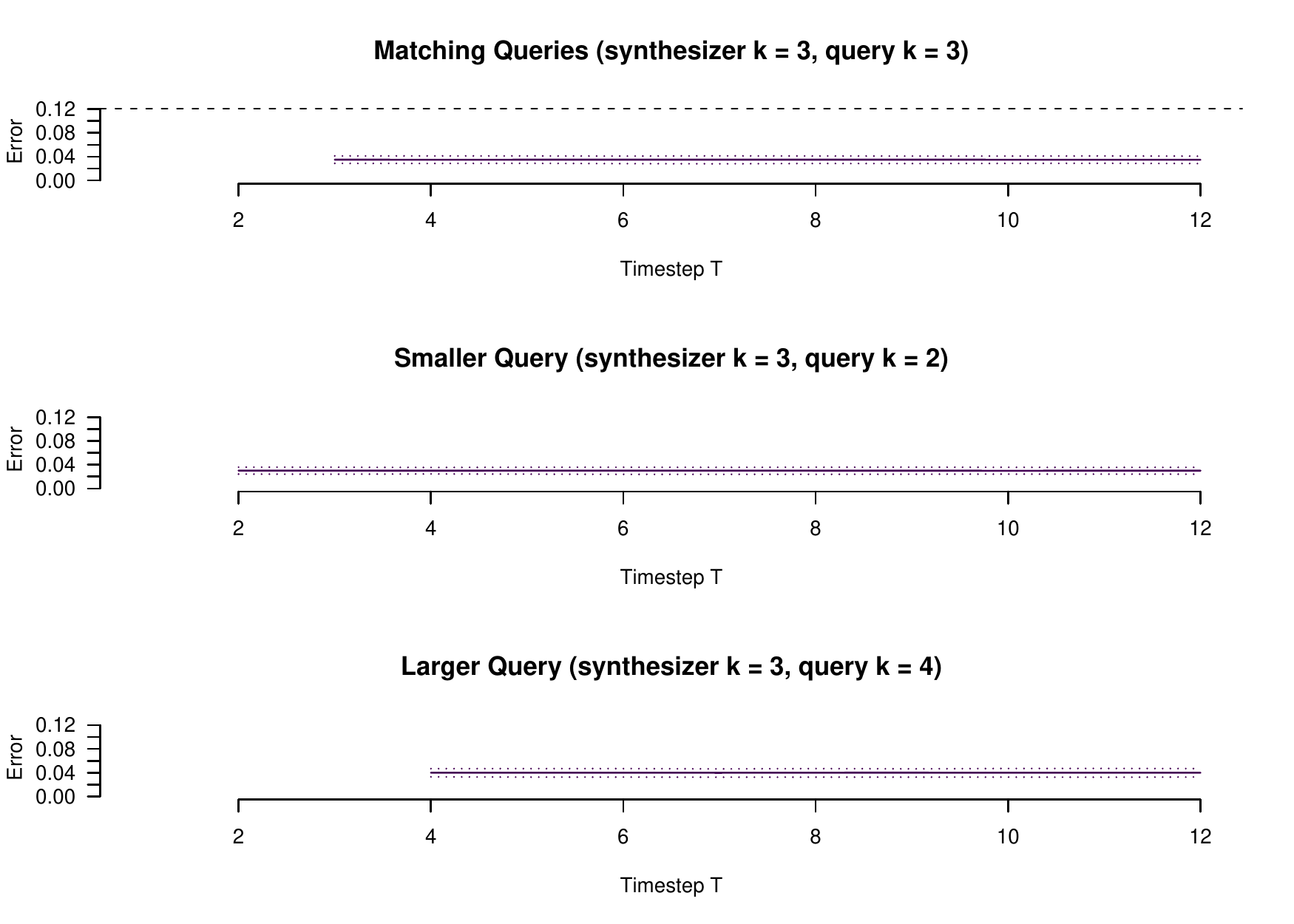}}
\caption{Empirical evaluation of the error of Algorithm \ref{algo:fixedwindow} on simulated data. The solid line shows the median error at each timestep across $1000$ repetitions of the algorithm. The dotted lines show the $2.5$ and $97.5$ percentile. The horizontal dashed line shows the theoretical error bound for proportions calculated on the synthetic data without the debiasing step. }
\label{fig:experiment12}
\end{center}
\vskip -0.2in
\end{figure}

In Figure \ref{fig:experiment12}, we show that the debiasing step is essential. Calculating the proportions on the synthetic data directly leads to a substantially larger error.

\subsection{Synthetic data for fixed time window queries}
\begin{figure}[ht]
\vskip 0.2in
\begin{center}
\centerline{\includegraphics[width=1\linewidth]{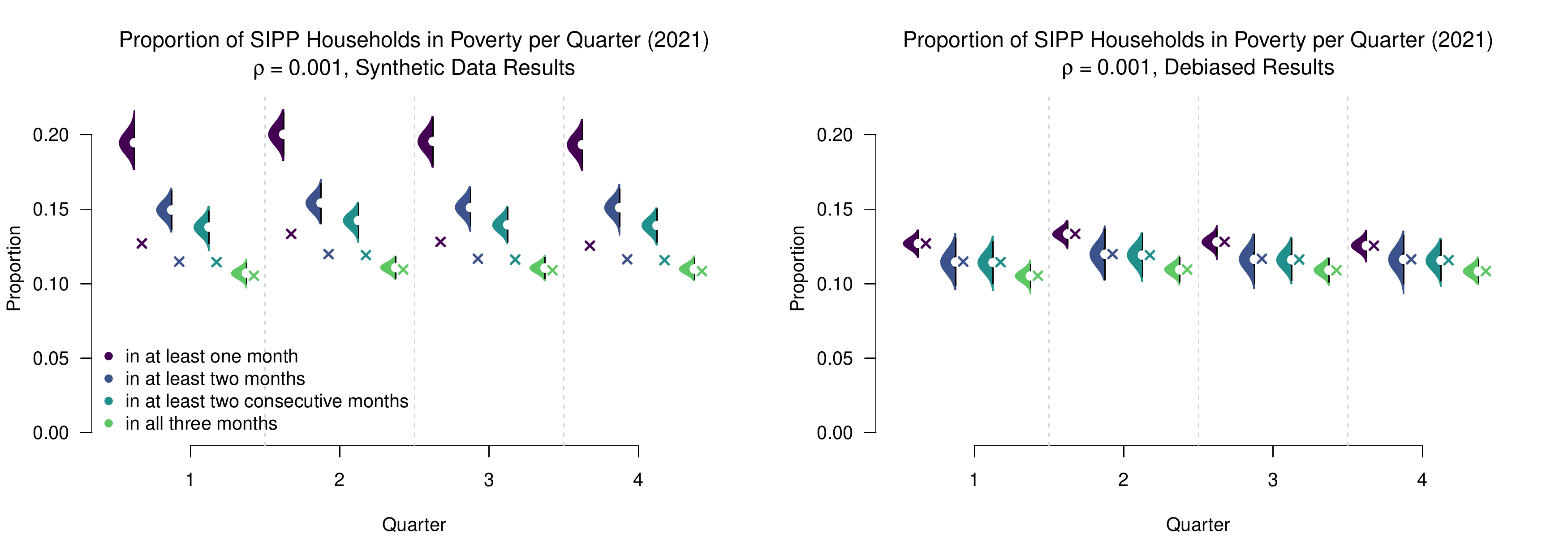}}
\caption{Left Panel: Proportions of SIPP Households in poverty per quarter in 2021. Calculated on the synthetic data, no debiasing step.  Right Panel: Proportions of SIPP Households in poverty per quarter in 2021. Calculated on the synthetic data, debiased by subtracting the result of the query run on the padding data.The density estimates show the empirical privacy noise distribution across $1000$ repetitions of the experiments with privacy parameter $\rho = 0.001$. X's indicate values calculated from the SIPP data.}
\label{fig-app:experiment32}
\end{center}
\vskip -0.2in
\end{figure}

\begin{figure}[ht]
\vskip 0.2in
\begin{center}
\centerline{\includegraphics[width=1\linewidth]{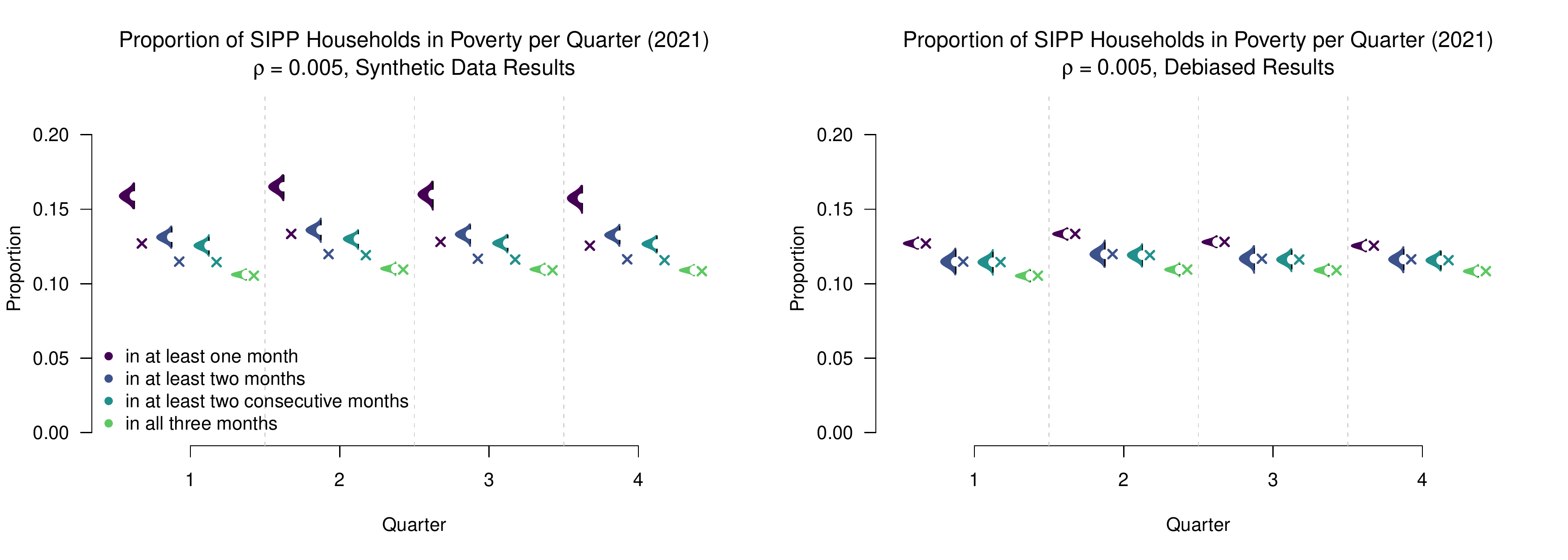}}
\caption{Left Panel: Proportions of SIPP Households in poverty per quarter in 2021. Calculated on the synthetic data, no debiasing step.  Right Panel: Proportions of SIPP Households in poverty per quarter in 2021. Calculated on the synthetic data, debiased by subtracting the result of the query run on the padding data.The density estimates show the empirical privacy noise distribution across $1000$ repetitions of the experiments with privacy parameter $\rho = 0.005$. X's indicate values calculated from the SIPP data.}
\label{fig-app:experiment31}
\end{center}
\vskip -0.2in
\end{figure}

\begin{figure}[ht]
\vskip 0.2in
\begin{center}
\centerline{\includegraphics[width=1\linewidth]{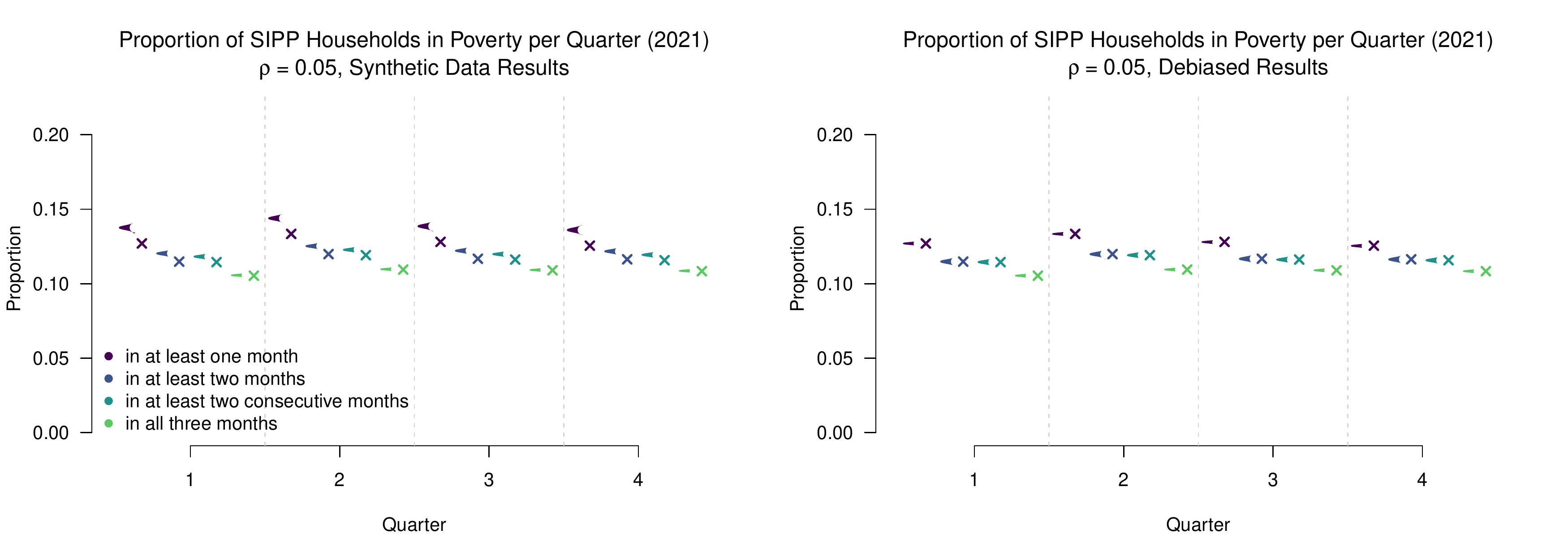}}
\caption{Left Panel: Proportions of SIPP Households in poverty per quarter in 2021. Calculated on the synthetic data, no debiasing step.  Right Panel: Proportions of SIPP Households in poverty per quarter in 2021. Calculated on the synthetic data, debiased by subtracting the result of the query run on the padding data.The density estimates show the empirical privacy noise distribution across $1000$ repetitions of the experiments with privacy parameter $\rho = 0.05$. X's indicate values calculated from the SIPP data.}
\label{fig-app:experiment33}
\end{center}
\vskip -0.2in
\end{figure}

In Figures \ref{fig-app:experiment32}-\ref{fig-app:experiment33} we present our results for different values of the privacy parameter $\rho$. In the right panels of Figures \ref{fig-app:experiment32}-\ref{fig-app:experiment33}, we show that our answers based on the synthetic data averaged over 1000 repetitions accurately match the ground truth, indicating that our approach provides an unbiased estimate of the fixed time queries (by subtracting the query answer on the padding data from the query answer on the complete synthetic data).

\subsection{Synthetic data based on cumulative time queries}

\begin{figure}[ht]
\vskip 0.2in
\begin{center}
\centerline{\includegraphics[width=0.5\linewidth]{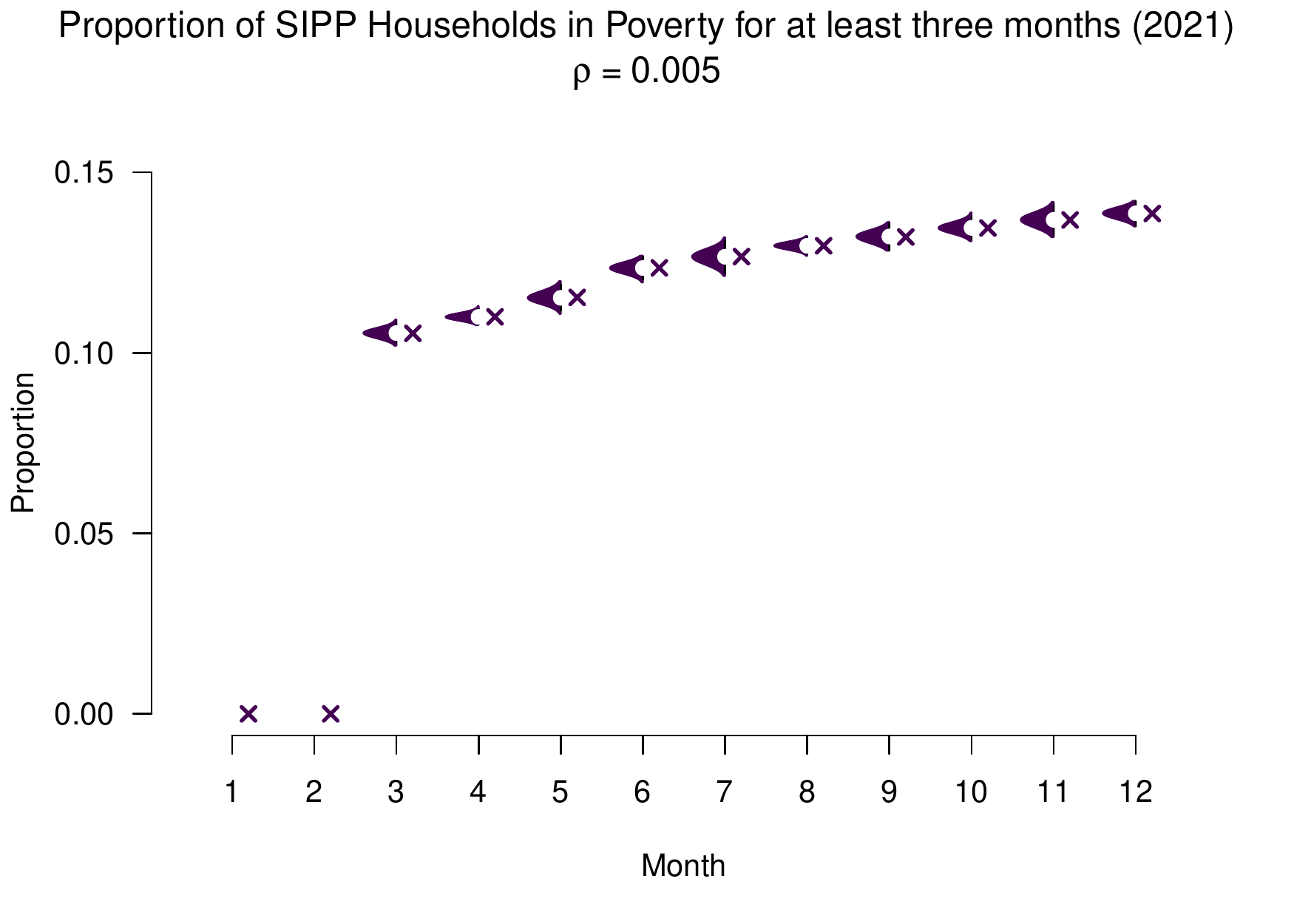}}
\caption{Proportion of SIPP Households in poverty for at least three months ($b = 3$) up to any given month in 2021. The density estimates show the empirical privacy noise distribution across $1000$ repetitions of the experiments with privacy parameter $\rho = 0.005$. X's indicate values calculated from the SIPP data.}
\label{fig-app:experiment4}
\end{center}
\vskip -0.2in
\end{figure}

In Figure \ref{fig-app:experiment4}, we again repeat the synthesizer 1000 times and plot the answers to these queries using our generated DP synthetic data versus the ground truth, with the x-axis denoting the time horizon. Our answers based on the synthetic data averaged over 1000 repetitions accurately match the ground truth, indicating that our approach provides an unbiased estimate of the cumulative time queries.

While Algorithm \ref{algo:cumulative} generates synthetic data for all time thresholds $b$ from $1, \dots, T$ simultaneously, we here focus on the results for setting the threshold to $b = 3$.

\end{document}